\documentclass[twocolumn]{aastex61}
\usepackage{amsmath}
\usepackage{lipsum}
\usepackage{graphics}
\usepackage{scrextend}
\usepackage{url}
\usepackage{breakurl}
\deffootnote{2em}{0em}{\thefootnotemark\quad}

\begin{document}

\title{AzTEC Survey of the Central Molecular Zone: Data Reduction, Analysis, and Preliminary Results}

\author{Yuping Tang}
\affiliation{Department of Astronomy, University of Massachusetts Amherst, Amherst, 01002, USA}
\affiliation{Chinese Academy of Sciences South America Center for Astronomy, National Astronomical Observatories, CAS, Beijing, 100012, China}

\author{Q. Daniel Wang, Grant W. Wilson, Mark H. Heyer, Robert A. Gutermuth, Peter Schloerb, Min S. Yun}
\affiliation{Department of Astronomy, University of Massachusetts Amherst, Amherst, 01002, USA}

\author{John Bally}
\affiliation{Center for Astrophysics and Space Astronomy, Astrophysical and Planetary Sciences Department, University of Colorado, UCB 389 Boulder, Colorado 80309, USA}

\author{Laurent Loinard}
\affiliation{Instituto de Radioastronom\'{\i}a y Astrof\'{\i}sica, Universidad Nacional Aut\'{o}noma de M\'{e}xico
Apartado Postal 3-72, 58090, Morelia, Michoac\'{a}n, Mexico}

\author{Sergiy Silich}
\affiliation{Instituto Nacional de Astrof\'{i}sica, \'{O}ptica y Electr\'{o}nica (INAOE), Luis Enrique Erro 1, Tonantzintla, Puebla, 72840, Mexico}

\author{Miguel Ch\'{a}vez}
\affiliation{Instituto Nacional de Astrof\'{i}sica, \'{O}ptica y Electr\'{o}nica (INAOE), Luis Enrique Erro 1, Tonantzintla, Puebla, 72840, Mexico}

\author{Daryl Haggard}
\affiliation{Department of Physics, McGill University, 3600 University Street, Montréal, QC H3A 2T8, Canada}

\author{Alfredo, Monta\~{n}a}
\affiliation{Consejo Nacional de Ciencia y Tecnolog\'{\i}a. Av. Insurgentes Sur 1582, 03940, Ciudad de M\'{e}xico, Mexico}
\affiliation{Instituto Nacional de Astrof\'{i}sica, \'{O}ptica y Electr\'{o}nica (INAOE), Luis Enrique Erro 1, Tonantzintla, Puebla, 72840, Mexico}

\author{David S\'{a}nchez-Arg\"{u}elles}
\affiliation{Instituto Nacional de Astrof\'{i}sica, \'{O}ptica y Electr\'{o}nica (INAOE), Luis Enrique Erro 1, Tonantzintla, Puebla, 72840, Mexico}

\author{Milagros Zeballos}
\affiliation{Instituto Nacional de Astrof\'{i}sica, \'{O}ptica y Electr\'{o}nica (INAOE), Luis Enrique Erro 1, Tonantzintla, Puebla, 72840, Mexico}
\affiliation{Universidad de las Am\'{e}ricas Puebla, Sta. Catarina M\'{a}rtir, San Andr\'{e}s Cholula, 72810 Puebla, Mexico}

\author{Jorge A. Zavala}
\affiliation{Instituto Nacional de Astrof\'{i}sica, \'{O}ptica y Electr\'{o}nica (INAOE), Luis Enrique Erro 1, Tonantzintla, Puebla, 72840, Mexico}
\affiliation{The University of Texas at Austin, 2515 Speedway Blvd Stop C1400, Austin, TX 78712, USA}

\author{Jonathan Le\'{o}n-Tavares}
\affiliation{Instituto Nacional de Astrof\'{i}sica, \'{O}ptica y Electr\'{o}nica (INAOE), Luis Enrique Erro 1, Tonantzintla, Puebla, 72840, Mexico}
\affiliation{Centre for Remote Sensing and Earth Observation Processes (TAP-Remote Sensing), Flemish Institute for Technological Research (VITO), Boeretang 282, B-2400 Mol, Belgium}

\begin{abstract}
We present a large-scale survey of the central molecular zone (CMZ) of our Galaxy, as well as a monitoring program of Sgr A*, with the AzTEC/Large Millimeter Telescope (LMT) in the 1.1 mm continuum. Our 1.1 mm map covers the main body of the CMZ over a field of $1.6 \times 1.1$ deg$^2$ with an angular resolution of $10.5''$ and a depth of 15 mJy/beam. To account for the intensity loss due to the background removal process, we combine this map with lower resolution CSO/Bolocam and \textit{Planck}/HFI data to produce an effective full intensity 1.1 mm continuum map. With this map and existing \textit{Herschel} surveys, we have carried out a comprehensive analysis of the spectral energy distribution (SED) of dust in the CMZ. A key component of this analysis is the implementation of a model-based deconvolution approach, incorporating the Point Spread Functions (PSFs) of the different instruments, and hence recovering a significant amount of spatial information on angular scales larger than $10.5''$. The monitoring of Sgr A* was carried out as part of a worldwide, multi-wavelength campaign when the so-called G2 object was undergoing the pericenter passage around the massive black hole (MBH). Our preliminary results include 1) high-resolution maps of column density, temperature and dust spectral index across the CMZ; 2) a 1.1~mm light curve of Sgr A* showing an outburst of $140\%$ maximum amplitude on 9th May, 2014 but otherwise only stochastic variations of $10\%$ and no systematic long-term change, consistent with other observations. 
\end{abstract}


\section{Introduction} \label{sec:intro}

\subsection{The Galactic Central Molecular Zone}

The Galactic Central Molecular Zone (CMZ) contains the highest concentration of (sub-)mm dense clouds, and $80\%$ of the dense molecular gas in the Galaxy~\citep{longmore13}. Some of these clouds have been identified as candidate young massive cluster (YMC) precursors ~\citep[e.g., Sgr B2 \& Sgr C]{ginsburg18, lu19}. These clouds may form star clusters, similar to the three well studied clusters in the Galactic Center (Arches, Quintuplet, and GC star cluster), which are a few $10^6$ yrs old. Yet, it has been realized for years that the \textit{global} star formation rate (SFR) in the CMZ is an order of magnitude lower than expected from empirical relationships established in both the local and high-z universe \citep{yusef09a, longmore13, kruijssen14, kauffmann17}. Most dense clouds and regions in the CMZ (e.g., G0.253+0.016, ``The Brick'') with $n_{peak}>10^{6}$ $cm^{-3}$) show weak or no ongoing star formation activity. This could be potentially attributed to supporting pressures against gravitational collapse from a variety of origins, including magnetic fields~\citep{pillai15}, cosmic rays~\citep{indriolo14,oka19},  turbulent motions~\citep{kruijssen13} and stellar feedback~\citep{armillotta19}. While it is suggested that the dominant suppressor of the star-formation is the turbulent pressure \citep{kruijssen14}, more detailed scenarios are being established in individual CMZ clouds ~\citep[e.g., G0.253+0.016]{federrath16}. New dynamical models propose that the star formation in the CMZ is episodic and currently in a slump, but may be triggered toward a burst phase, possibly by tidal compression induced by the Galactic gravitational potential, as gas accumulation due to bar-driven inflows reaches a threshold \citep{kruijssen15, kruijssen19}. In this picture, dense clouds in the CMZ could geometrically form a star-forming sequence, where star formation in each cloud is triggered during its pericentre passage while most clouds are currently in a pre-burst phase.  

In this paper, we will present a map of the CMZ obtained with the AzTEC camera \citep{wilson08} mounted on the LMT, which provides the first high-resolution ($10.5''$ beam) continuum survey of the field at 1.1 mm. At this wavelength, the emission is optically thin even for YMC precursors. Furthermore, being squarely in the Raleigh-Jeans portion of the dust SED, the 1.1 mm flux is only  linearly dependent on dust temperature, making it a promising tracer of total gas column densities. LMT/AzTEC imaging, together with existing complementary surveys such as those from \textit{Herschel} at substantially shorter wavelengths ($160\mu m$ for comparable resolutions), enables us to detect YMC precursors and other dense clouds with little bias. It is then possible to characterize their structural and physical properties, as well as the size and mass distributions, and to probe how such clouds form in the GC environment and under what conditions SF takes place. The linear resolution and sensitivity of the LMT/AzTEC GC mapping improve those of earlier studies at comparable wavelengths): the BOLOCAM/CSO survey at 1.1 mm ~\citep[30 mJy per $33''$ beam]{ginsburg13} the LABOCA/APEX survey at $850\mu m$ ~\citep[80mJy per $19''$ beam]{csengeri16} and the SCUBA2/JCMT survey at $850\mu m$ ~\citep[43mJy per $13''$ beam]{parsons18}). AzTEC/LMT provides the missing high-resolution coverage of dust emission from the CMZ in a long wavelength band. With this capability, we can spatially resolve cloud structures down to scales of 0.4~pc. Interestingly, recent interferometer observations of the CMZ clouds suggest a critical spatial scale of $0.1-1$ pc: For scales larger than $\approx 1$ pc, the CMZ clouds are distinct from the ``classic'' Milky Way (MW) disk clouds in terms of their smooth density structure ~\citep[e.g., G0.253+0.016]{rathborne15} and high-velocity dispersion at a given size ~\citep{kauffmann17}. Below $0.1-1$ pc, the clouds in the CMZ are similar to those in the Galactic disk in terms of their core to star–formation efficiencies ~\citep{lu19}. 

To maximize the utilization of a high-resolution map at 1.1 mm and to achieve optimal spatial resolution for our SED analysis, a proper treatment of multi-wavelength, multi-resolution maps is needed. The conventional approach in this case is to dilute all maps to a common low resolution, which results in the loss of information. Alternatively, the dilution of the Point Source Function (PSF) could be incorporated as an additional ingredient of the model in a forward-fitting manner. However, direct convolution of a large map with its PSF is computationally expensive. 

In this study, we develop a
computationally-economic, forward-fitting approach to solve resolution non-uniformity among different maps.  With this SED analysis, we can pursue the following 
objectives: 1) to unbiasedly and systematically characterize dense clouds in the CMZ by their temperature and density structure; and 2) to investigate how dust properties vary with gas density and temperature in this extreme environment that shares similarities with environments in high-redshift starburst galaxies. 

\subsection{Activity of Sgr A*}

We also present in this paper a monitoring program of Sgr A* conducted simultaneously with the large-scale survey, in an effort to characterize the light-curve of Sgr A* when the so-called G2 object was undergoing pericenter passage around the massive black hole (MBH) in 2014. The monitoring program was naturally integrated with the mapping program since Sgr A* also served as a pointing source for the former. The monitoring data was obtained right before or after the survey observations on a given night.  

The MBH at the GC is associated with the compact radio source, Sgr A*. 
Sgr A* is usually described by various classes of radiatively inefficient accretion flows (RIAFs), but is also an “underfed” MBH~\citep{narayan02, yuan03, wang13}. 
The thermal plasma emission between $10^4-10^5$ R$_S$ is responsible for the bulk of the quiescent X-ray emission observed from Sgr A*, where R$_S$ is the Schwarzschild radius of the BH. By contrast, X-ray flares at a rate of $\sim 2$ per day and on time scales of $\sim 1$ hr~\citep{yuan18, mossoux20}, as well as highly variable near-infrared (NIR) emission\citep{eckart12,neilsen13,gravity18}. 
Such flares and other variability behaviors of Sgr A*~\citep{li09} are poorly understood, although several scenarios have been widely explored, such as hot spots in the accretion flow and expanding hot bubbles ~\citep[probably due to magnetic field reconnection or other accretion disk instabilities]{gravity20}. The flat-spectrum radio emission of Sgr A* is typical for radio cores in jet sources and under-predicted in many accretion models. Therefore, jets~\citep{falcke00a} or jet-RIAF hybrids~\citep{yuan02} are often invoked; but none of these models are well tested or constrained. The understanding of the emission and its variability is not only crucial for understanding how the ``silent majority'' of SMBHs work, but also for attempts to image the event horizon of Sgr~A* \citep{falcke00b, broderick06}. 

The discovery of a highly reddened Br-gamma-emitting object (G2), dispersing a significant amount of matter while having a close encounter with Sgr A*, provided an excellent opportunity to study fundamental accretion physics~\citep{pfuhl15}. According to simulations and observations, the cloud passed its pericenter in the middle of 2014 (with considerable uncertainty) at a distance of only $2000$ R$_S$ from the BH~\citep{gillessen13}. The predicted average feeding rate during the encounter, consistent with existing observations of spatially and kinematically resolved tidal streams of gas, is $\sim10^{-7}M_\odot$/yr, comparable to, or higher than, the quiescent accretion rate in the innermost region of Sgr A*~\citep{burkert12,ballone13}. 

The present quiescent state of Sgr A* suggests an accretion rate that is much smaller than predicted by the Bondi model. Occasional giant outbursts of Sgr A* may make it substantially brighter. These could also shape the surrounding medium on various scales, such as X-ray reflection or Fermi Bubbles. The impact of such outbursts could play essential roles in the structural and thermal properties of the gas in the nuclear region. \bigskip 

The structure of this paper is organized as follows. The observation strategy and data reduction for our AzTEC survey are described in Section~\ref{sec:obs} and Section~\ref{sec:redu}, respectively. Section~\ref{sec:redu} also describes how \textit{Herschel}-SPIRE, \textit{Planck}-HFI, and CSO-Bolocam observations are processed and analyzed in our study. In Section~\ref{sec:deconvol}, we describe and demonstrate a model-based deconvolution approach for SED analysis. In Section~\ref{sec:results}, we present high-resolution maps of $N_{H_2}$, $T$ and $\beta$ and also the lightcurve of the Sgr A*. In Section~\ref{sec:summary} we draw our conclusions. Throughout this work, the distance to the Galactic center is assumed to be 8.5 kpc.

\section{Observations}\label{sec:obs}

\subsection{Mapping}\label{subsec:mapping}

\begin{figure*}
\begin{center}
\includegraphics[scale=0.35]{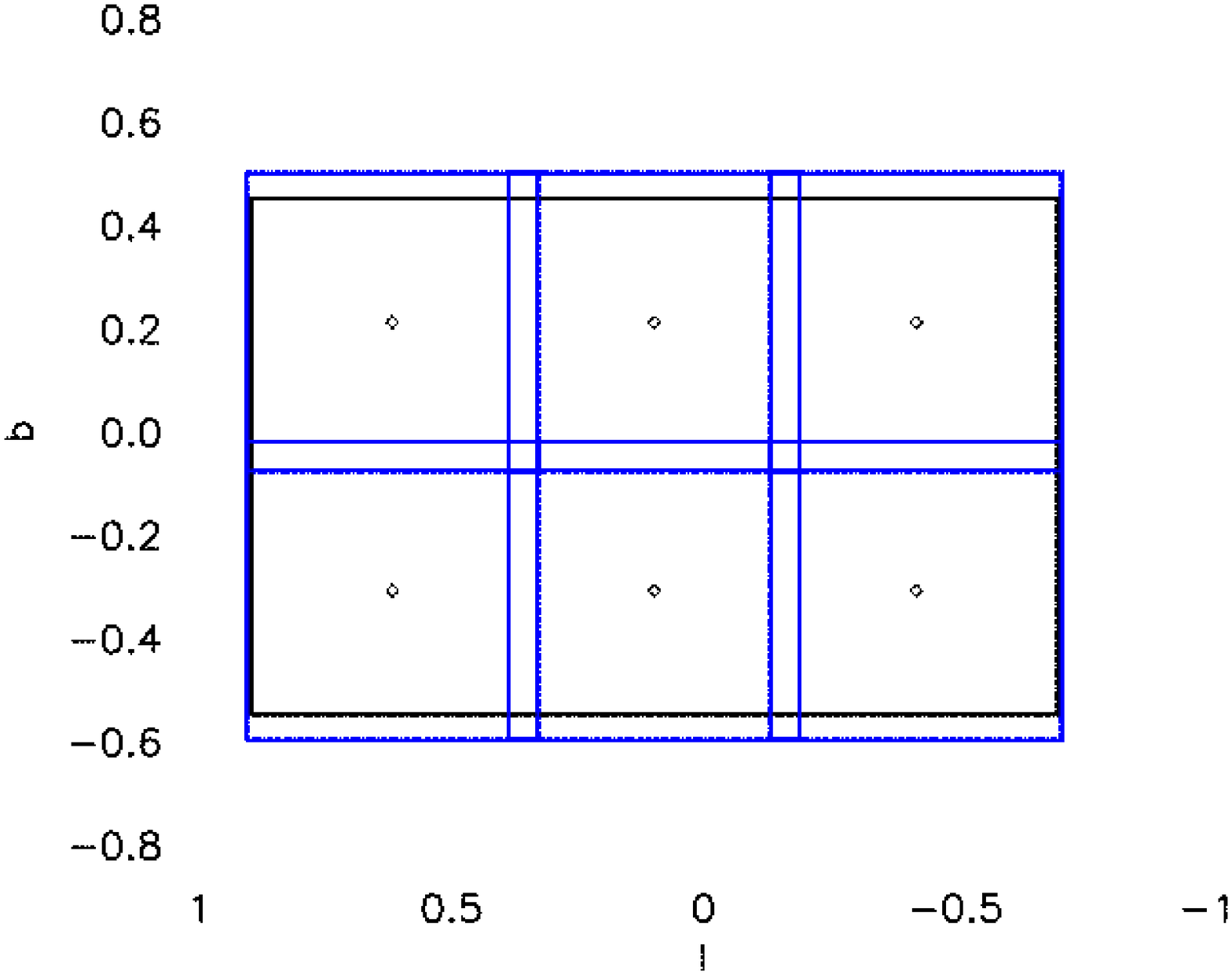}
\includegraphics[scale=0.35]{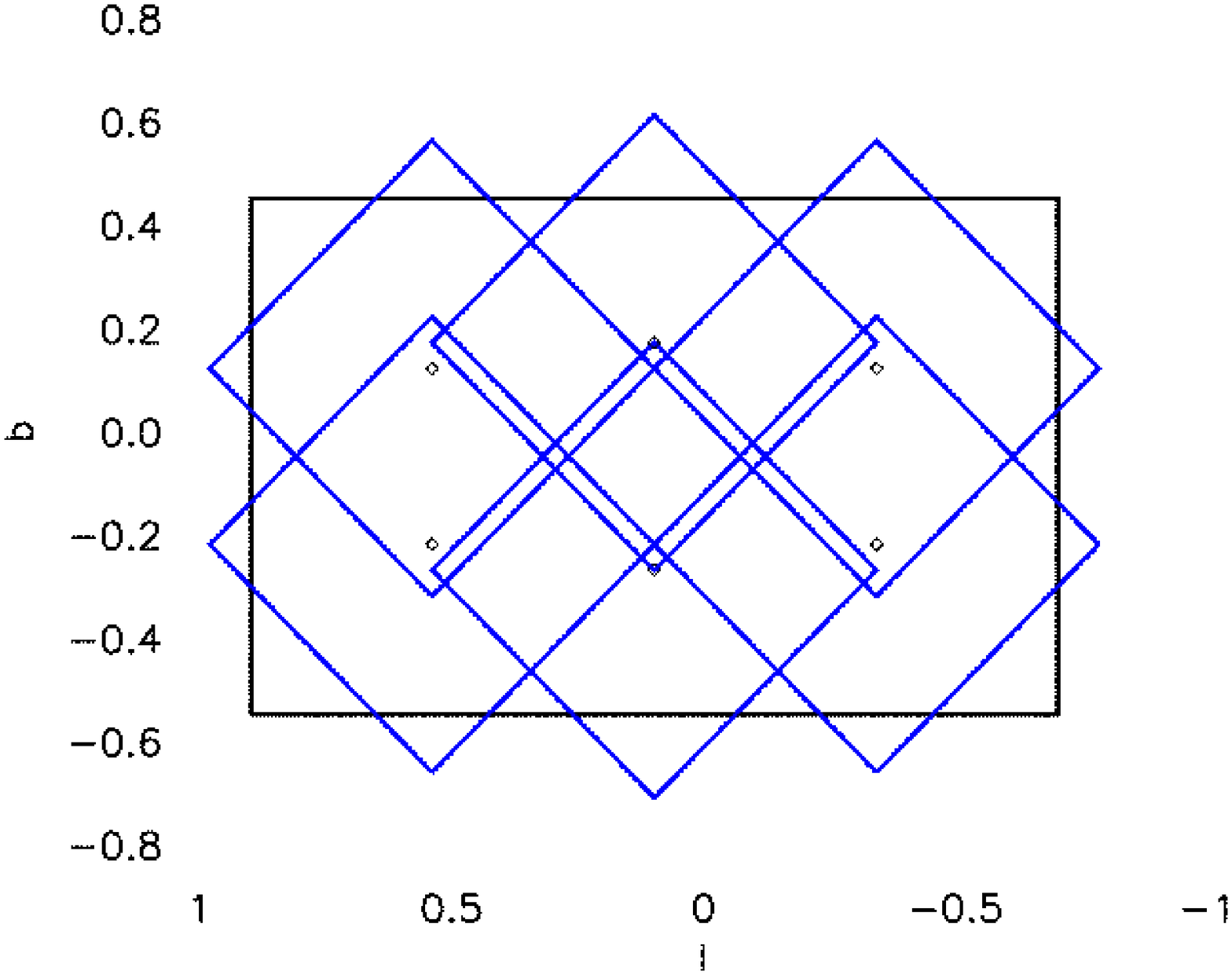}
\end{center}
\caption{Two different patterns of tiles covering the target Galactic Center Region: $l=[-0.7^{\circ}, 0.9^{\circ}]$, $b=[-0.6^{\circ}, 0.5^{\circ}]$. In each pattern, every tile has an identical size. The first pattern has a tile size of $\approx 0.57^{\circ}$, and the second pattern has a tile size of $\approx 0.62^{\circ}$.
}
\label{fig:tiles}
\end{figure*}

Our AzTEC survey was conducted during the Early Science 2 (ES2) campaign of the 32 meter LMT in 2014. The AzTEC bolometer array has 113 functioning detectors, with a field of view (FOV) $=2.4  \times 2.4$ arcmin$^2$. The survey covers the Galactic Center Region $l=[-0.7^{\circ}, 0.9^{\circ}]$, $b=[-0.6^{\circ}, 0.5^{\circ}]$, which roughly extends from Sgr B2 to Sgr C. The target field was mosaiced by 6 square tiles in 2 different patterns, as shown in Figure~\ref{fig:tiles}. This survey strategy is adopted for two reasons. First, the CMZ has a maximum elevation of $\approx 43^\circ$ as seen from the LMT located at the summit of Sierra Negra, Mexico, at an elevation of 4600 meters and geographic latitude of 18$^\circ$. During ES2, the telescope gain was a non-negligible function of the altitude. 
By dividing our target field into small tiles, each tile can be calibrated with a roughly constant gain factor. %

Each tile was observed with a raster-scan mode, meaning that a square tile is scanned line by line, \textbf{back-and-forth}, with a scanning speed of $200''/s$ and a step size of $30''$ perpendicular to the scanning direction. The observations were  carried out under average to good weather conditions ($\tau_{1.1mm} \lesssim 0.2$). The parameters of all observed tiles are summarized in Table~\ref{tab:tiles}. Each pattern (with 6 tiles) takes $\approx 2$ hours to complete. In practice, we scanned the entire field once with either Pattern 1 or 2 during each night of observation, along with pointing observations. We use Sgr A* and a compact millimeter source BGPS G$000.378+00.041$ as pointing sources. Science observations were interwoven with pointing observations between every two tiles for corrective pointing and focus offsets of the secondary mirror. The survey was conducted from Apr 17 to June 18, 2014, with a total integration time of $\approx 20$ hours and $9$ individual maps, evenly distributed between Pattern 1 and Pattern 2.

\begin{figure}
\includegraphics[scale=0.4]{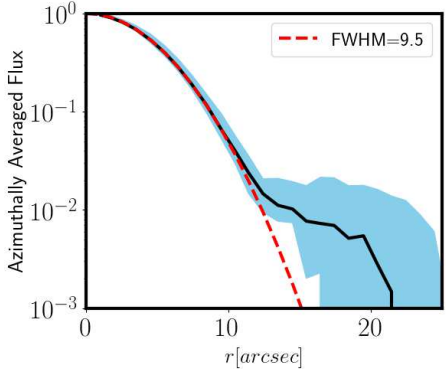}

\caption{The azimuthally averaged radial profile (solid black) of the AzTEC PSF. The PSF is derived from an observation of Ceres in ``beammap'' mode, at an elevation of 40''. $\pm 1 \sigma$ errs are indicated by the blue shade. The primary lobe is well fitted with a Gaussian (red-dotted, FWHM=9.5''). The response of the first side lobe, $\approx-20$ dB, is in agreement with \cite{wilson08}.}
\label{fig:ceres}
\end{figure}

The altitude of the CMZ is low ($b=37-42^{\circ}$) as seen from the LMT. This is close to the lower-limit of altitude where the telescope could operate within reasonable pointing uncertainties during ES2. As a result, the effective beam size area is larger than the standard value ($8.5''$) for the 32 meter LMT. We have conducted beam mapping observations toward asteroids, these observations were used to determine the intrinsic beam size during the observation. We find an intrinsic beam size of $9.5 \pm 0.5''$(Figure~\ref{fig:ceres}). This value is used for beam match calibration with other sets of data, i.e., conversion from Jy/beam to Jy/sr.

\begin{figure}
\includegraphics[scale=0.4]{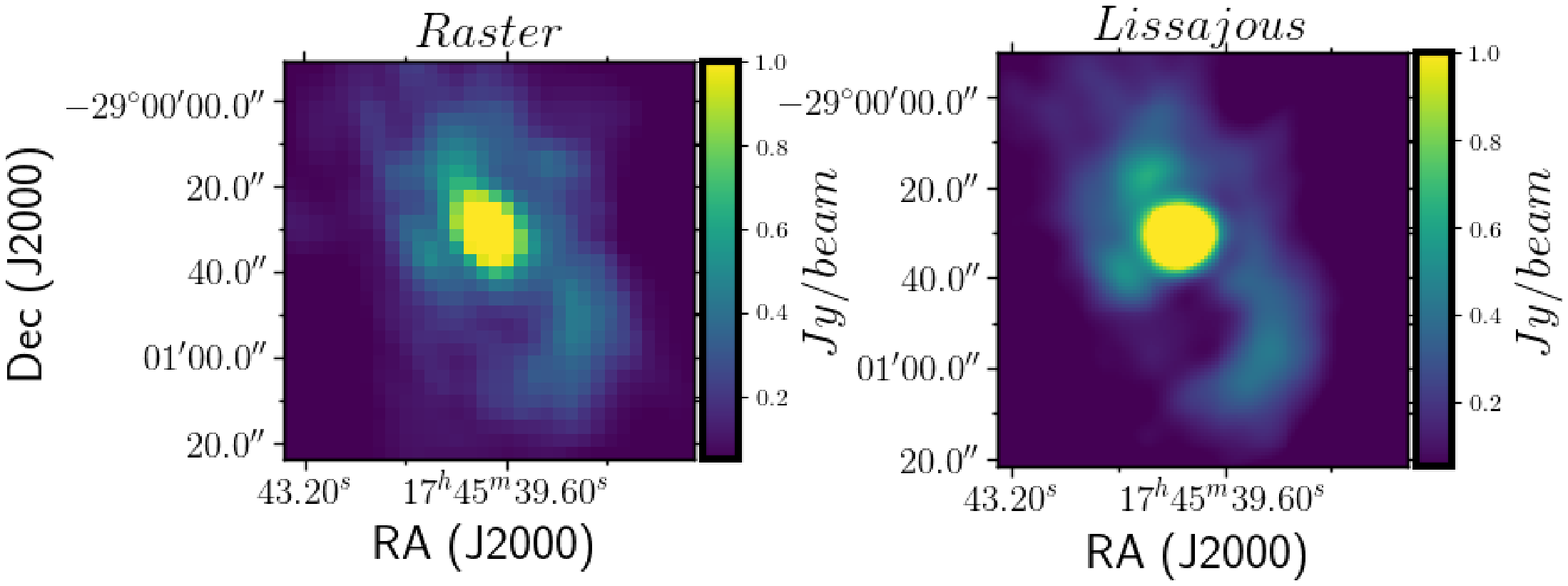}

\caption{A comparison between a fast scanned raster map (pixel size = $3''$) and a slowly scanned Lissajous map (pixel size = $0.75''$).
Note that the sampling rates are different ($v_{scan} = 50''/s$ for the Lissiajous map 
and $v_{scan} \approx 200''/s$ for the raster map). The elongation of beam shape along the scanning direction could be clearly seen in the raster map. In this single case, the beam size along the scanning direction is elongated to $10.5''$, which is consistent with our simulation.}
\label{fig:beam_shape}
\end{figure}

We compare the raster maps with the pointing observations towards Sgr A*. The later ones are obtained with a Lissajous scanning pattern with a lower scanning speed, $50''/s$, as shown in Figure~\ref{fig:beam_shape}.  The elongated beam size along the scanning direction is $10.5''$ on average. Since every pixel is scanned along 8 different directions, we use $10.5''$ as a rough estimate for the effective resolution for the product map. Notice that this larger $10.5''$ beam size is due to under-sampling, which differs from the beam size used for flux calibration (unit conversion from Jy/beam to Jy/sr). The latter is determined from pointing observations, which is carried out with slow scanning speed, thus unaffected by under-sampling, and is found to be $\approx 9.5''$, as mentioned before.

\begin{table*}[]
\centering
\caption{List of AzTEC tiles covering the CMZ}
\label{tab:tiles}
\resizebox{\textwidth}{!}{
\begin{tabular}{cccccccccccc}
 Observation ID & RA & DEC & Pattern & Dimension & Date & Observation ID & RA & DEC & Pattern & Dimension & Date \\
  \hline
 019329 & 17:45:49.85  & -29:27:20.67 & A & 2080'' & 04/17/2014 & 021854 & 17:47:04.09  & -29:00:42.33 & A & 2080'' & 06/02/2014 \\
 019330 & 17:47:04.09  & -29:00:42.33 & A & 2080'' & 04/17/2014 & 021855 & 17:45:49.85  & -29:27:20.67 & A & 2080'' & 06/02/2014 \\
 019331 & 17:48:17.71  & -28:34:01.51 & A & 2080'' & 04/17/2014 & 021858 & 17:43:47.89  & -29:11:03.30 & A & 2080'' & 06/02/2014\\
 019334 & 17:46:16.41  & -28:17:52.43 & A & 2080'' & 04/17/2014 & 021859 & 17:45:02.47  & -28:44:29.15 & A & 2080'' & 06/02/2014 \\
 019335 & 17:45:02.47  & -28:44:29.15 & A & 2080'' & 04/17/2014 & 021860 & 17:46:16.41  & -28:17:52.43 & A & 2080'' & 06/02/2014 \\
 019336 & 17:43:47.89  & -29:11:03.30 & A & 2080'' & 04/17/2014 & 022762 & 17:47:45.70  & -28:35:17.77 & B & 2245'' & 06/15/2014 \\
 019404 & 17:43:47.89  & -29:11:03.30 & A & 2080'' & 04/20/2014 & 022763 & 17:46:54.75  & -28:59:28.02 & B & 2245'' & 06/15/2014 \\
 019405 & 17:45:02.47  & -28:44:29.15 & A & 2080'' & 04/20/2014 & 022764 & 17:45:40.13  & -29:20:31.46 & B & 2245'' & 06/15/2014 \\
 019411 & 17:46:16.41  & -28:17:52.43 & A & 2080'' & 04/20/2014 & 022767 & 17:45:11.76  & -28:45:43.94 & B & 2245'' & 06/15/2014\\
 019412 & 17:45:49.85  & -29:27:20.67 & A & 2080'' & 04/20/2014 & 022769 & 17:45:11.76  & -28:45:43.94 & B & 2245'' & 06/15/2014 \\
 019413 & 17:47:04.09  & -29:00:42.33 & A & 2080'' & 04/20/2014 & 022848 & 17:44:20.14  & -29:09:50.56 & B & 2245'' & 06/16/2014 \\
 019478 & 17:47:45.70  & -28:35:17.77 & B & 2245''  & 04/21/2014 &  022850 & 17:45:11.76  & -28:45:43.94 & B & 2245'' & 06/16/2014  \\
 019479 & 17:46:54.75  & -28:59:28.02 & B & 2245''  & 04/21/2014 & 022853 & 17:46:26.07  & -28:24:41.47 & B & 2245'' & 06/16/2014 \\
 019480 & 17:45:40.13  & -29:20:31.46 & B & 2245''  & 04/21/2014 & 022854 & 17:45:40.13  & -29:20:31.46 & B & 2245'' & 06/16/2014 \\
 019483 & 17:46:26.07  & -28:24:41.47 & B & 2245''  & 04/21/2014 & 022855 & 17:46:54.75  & -28:59:28.02 & B & 2245'' & 06/16/2014  \\
 019484 & 17:45:11.76  & -28:45:43.94 & B & 2245''  & 04/21/2014 & 022923 & 17:45:49.85  & -29:27:20.67 & A & 2080'' & 06/17/2014 \\
 019485 & 17:44:20.14  & -29:09:50.56 & B & 2245''  & 04/21/2014 & 022924 & 17:47:04.09  & -29:00:42.33 & A & 2080'' & 06/17/2014\\
 019489 & 17:48:17.71  & -28:34:01.51 & A & 2080'' &  04/21/2014 & 022927 & 17:48:17.71  & -28:34:01.51 & A & 2080'' & 06/17/2014\\
 021016 & 17:44:20.14  & -29:09:50.56 & B & 2245''  & 05/19/2014 & 022928 & 17:46:16.41  & -28:17:52.43 & A & 2080'' & 06/17/2014\\
 021017 & 17:45:11.76  & -28:45:43.94 & B & 2245''  & 05/19/2014 & 022929 & 17:45:02.47  & -28:44:29.15 & A & 2080'' & 06/17/2014\\
 021018 & 17:46:26.07  & -28:24:41.47 & B & 2245''  & 05/19/2014 & 023024 & 17:45:02.47  & -28:44:29.15 & A & 2080'' & 06/18/2014 \\
 021021 & 17:45:40.13  & -29:20:31.46 & B & 2245''  & 05/19/2014 & 023025 & 17:43:47.89  & -29:11:03.30 & A & 2080'' & 06/18/2014 \\
 021022 & 17:46:54.75  & -28:59:28.02 & B & 2245''  & 05/19/2014 & 023026 & 17:47:04.09  & -29:00:42.33 & A & 2080'' & 06/18/2014 \\
 021023 & 17:47:45.70  & -28:35:17.77 & B & 2245''  & 05/19/2014 & 023029 & 17:45:49.85  & -29:27:20.67 & A & 2080'' & 06/18/2014\\
 021853 & 17:48:17.71  & -28:34:01.51 & A & 2080'' &  06/02/2014 & 023030 & 17:46:16.41  & -28:17:52.43 & A & 2080'' & 06/18/2014\\
 \hline
\end{tabular}
}
\end{table*}

Another issue with our observations is associated with decreasing telescope gain at low altitudes. Flux calibrations with AzTEC images are performed on the pipeline level. A certain type of observation (named ``Beammap'') for calibration purposes toward planet/asteroid calibrators is performed about 2-3 times per night of AzTEC observation. At the end of each season, these Beammaps are collected to build a calibration curve to be implemented in the pipeline. However, during ES2, we don't have enough Beammaps observed at low elevations ($\approx 40^\circ$), mainly due to limited time slots toward the end of the night when the CMZ was observed. As a result, the telescope gain is overestimated because it is mainly determined by extrapolation from calibrators observed at higher altitudes. This problem is also evident in systematically decreasing flux in the pointing maps towards lower altitudes. As shown in Section~\ref{sec:res_monitor}, our monitoring program suggests that the telescope gain could have a $20-30\%$ uncertainty for each individual observation. 

We eventually obtain a $1\sigma$ noise level of about 15 mJy/beam after 20 hours integration. This is significantly worse than our target of 5 mJy/beam, due to poor weather conditions and partially due to loss of sensitivities at low altitudes during ES2. 

\subsection{Monitoring of Sgr A*} 

Sgr A* is observed as a pointing source (for first-order pointing correction) at a cadence of $\approx 2-3$ times each night of our survey, at the beginning, in the middle, and at the end of the set of CMZ observations. We used as calibrators two dense molecular clumps: BGPS G$000.378+00.041$ and AGAL G$359.984-00.009$.
These compact submillimeter sources are close to Sgr A* and are expected to have constant fluxes in contrast to quasars, commonly used as pointing sources in mm-astronomy. All the three sources are scanned in identical Lissajous patterns, with a speed of $v_{scan} = 50''/s$, and an integration time of 2 minutes over a region of $\approx 4 \times 4$ arcmin$^2$. During the nights of 9th May and 19th May, we also carried out relatively dense monitoring to sample short time scale variations over a period of 0.8 and 2.5 hours, respectively.

\section{Data Reduction}\label{sec:redu}

\subsection{Iterative Cleaning Based on Principle Component Analysis}

The raw data were reduced using the standard AzTEC analysis pipeline \citep{scott08}. We use iterative Principal Component Analysis (PCA) to remove correlated signals among bolometers, which are primarily contributed by atmospheric emissions, emissions from the telescope itself, and non-Gaussian noise associated with the secondary mirror and back-end instruments. Signal components projected onto the highest $N_e$ ranked eigenvectors in bolometer-bolometer space are viewed as non-astronomical signals and are removed, where $N_e$ is a user-defined integer. Since real astronomical signals corresponding to extended structures are also correlated at a certain level, indistinguishable from non-astronomical correlated signals in the eigenspace of bolometer arrays, a more aggressive (larger $N_e$) PCA cleaning removes both more non-astronomical and astronomical signals. To compensate for this loss of information, our cleaning is performed iteratively, until a conversion is reached such that the root-mean-square of values (rms) in a final noise map are consistent with no significant astronomical signal. This procedure has undergone systematic study when applied to the Bolocam data \citep{enoch06, ginsburg13} and is outlined in \cite{liu10} when applied to the AzTEC data. It comprises the following steps: \\
\\
1) PCA cleaning. The highest $N_e$  ranked eigen-components of signals are removed. A cleaned map with a pixel size of $3''$ is created. \\
\\
2) The cleaned map is slightly smoothed with a Gaussian kernel of FWHM $=3''$. Pixels above $3 \sigma$ significance in the smoothed cleaned map are identified and are preserved. Here, the noise level is predicted from bolometer sensitivities, which are modeled based on a set of observations toward planet and asteroid calibrators throughout the season. In practice, we find that jackknifed noise realizations (constructed by multiplying each time stream by $\pm1$ to suppress astronomical signals, see Section \ref{sec:noise}.) indicate systematically higher noise, within $20\%$ of the predicted noise level. This suggests that the S/N level we used
in practice was between 2.7 and 3.0$\sigma$. The jackknifed noise realization provides more conservative estimates. However, this realization was not directly implemented on the pipeline level for the iterative PCA cleaning. Hence we use noise predicted from bolometer sensitivities for this particular reduction procedure. Note that in post-reduction analyses we still use noise estimates from jackknifed noise realization.\\
\\
3) Every pixel in the cleaned map generated in Step 1) (unsmoothed) is zero-valued except $>3 \sigma$ pixels identified in Step 2) and also their neighboring pixels within an aperture of $9.5''$. This map is named ``\textbf{total map}''.\\ 
\\
4) The \textbf{total map} is cast back to the time-space and is subtracted from the raw data, the residuals are PC-removed again and
gridded to create a ``\textbf{residual map}'', which is then added to the current \textbf{total map}. This latter product is named ``\textbf{current map}''. \\
\\
5) Any new pixels above $3 \sigma$ in the \textbf{residual map} are preserved and are added to the \textbf{total map}.\\
\\
6) If at Step 5), no new pixels are found to be above $3 \sigma$ threshold, convergence is assumed to be reached, and the \textbf{current map} is our final product map. Otherwise, move back to Step 4).\\

We adopt $N_e = 4$. Eventually, a total of 20 iterations are performed until convergence. We found that taking $N_e=3-7$ lead to $<5\%$ differences in final recovered intensity.\\

\begin{figure*}
\begin{center}
\includegraphics[scale=0.4]{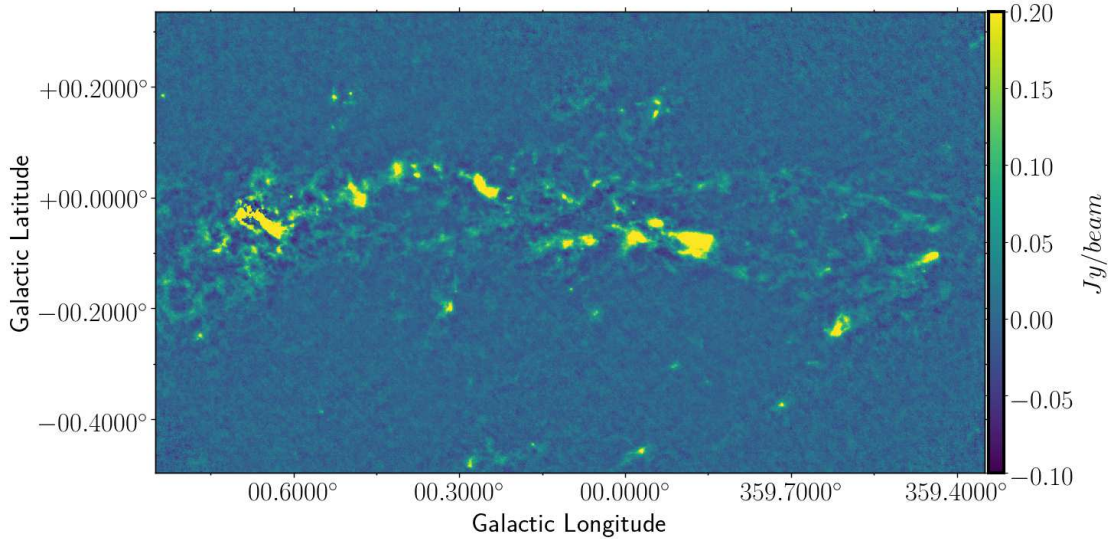}
\end{center}
\caption{Product AzTEC map of the CMZ. }
\label{fig:raw_map}
\end{figure*}

The product AzTEC map is shown in Figure~\ref{fig:raw_map}, which has a pixel size of $3''$. The region immediately around Sgr B2 is excluded from this analysis due to Sgr B2's brightness, which exceeds 10-15 Jy and is subsequently not treated properly by our analysis due to the extreme dynamic range of flux and the possibility that the detector response is non-linear for the source.

\subsection{Noise Maps}\label{sec:noise}

During the last iteration of PCA cleaning, a noise map is generated from cleaned residual signals. Typically, there is still a small, but visible, fraction of un-recovered extended emission in this residual map. These features can be further suppressed by multiplying each scan of residual signals in time-space (here, a scan is defined as all samples between any two adjacent turning points in a raster pattern) by randomly $+1$ or $-1$. This procedure further removes middle-to-large correlated signals in the scan, which survive PCA cleaning. The local standard deviation for each pixel is then calculated within a box of $\pm1'$ centered at it. The standard deviation derived in such a way is systematically higher than, but within $120\%$ of, what is predicted from bolometer sensitivities, which is not unexpected since we use a less aggressive PCA cleaning for our map compared to what we used to estimate bolometer sensitivities, therefore retaining more correlated noise corresponding to lower rank eigen-components. The varying and high scanning speed throughout this set of observations should also contribute to the deviation from prediction.

\subsection{Correction for Pointing offsets between Individual Tiles}
Pointing offsets are corrected on two levels, a pipeline level, and a post-pipeline level. On the pipeline level, pointing offsets are estimated from the most recent pointing observation (for this project, it is either toward Sgr A* or toward BGPS G$000.378+00.041$). On the post-pipeline level, each individual raster map is PCA cleaned and reduced separately. The optimal pointing offsets and zero-flux offsets are obtained by minimizing connecting differences among individual tiles. More specifically, we are minimizing:

\begin{equation}\label{eq:cross}
\chi^2 = \sum\limits_{i,j} \frac{\textbf{off}_{(i)} - \textbf{off}_{(j)}  + \textbf{diff}_{(ij)}}{\sigma_{(ij)}}
\end{equation}

Where $\textbf{diff}_{(ij)}$ stands for relative offset in either position or flux between two adjacent maps, the $i$th and the $j$th map. $\textbf{off}_{(i)}$ and $\textbf{off}_{(j)}$ are the absolute offsets to be solved. This is achieved by solving:
\begin{equation}
\frac{\partial \chi^2}{\partial \textbf{off}_{(i)}} = 0
\end{equation}

A more detailed description of this approach can be found in \cite{dong11} (Appendix A).

For pointing offsets, each $\textbf{diff}_{(ij)}$ in eq~\ref{eq:cross} is derived by cross-correlating a pair of two adjacent raster maps (i,j). Before cross-correlation, both maps are interpolated to a common grid with a pixel scale of $1''$, pixels below $3 \sigma$ significance are zero-valued to exclude correlated noise. The cross-correlated image is convolved with a Gaussian kernel of FWHM $= 10.5''$ prior to peak-finding. Following \cite{tonry79} and \cite{kurtz98}, we set the uncertainty $\sigma_{(ij)}$ of the cross-correlated peak to $\frac{3w}{8(1+r)}$, where $w$ is the FWHM of the correlation peak, and $r$ is the ratio of the correlation peak height to the amplitude of the noise.

To correct for relative offsets in zero-flux, similarly, two adjacent maps (i,j) are first interpolated to a common grid. In this case, $\textbf{diff}_{(ij)}$ is the difference of flux in shared pixels in the re-gridded images of map i and map j. $\sigma_{(ij)}$ is proportional to Poisson noise, $\sqrt[]{n_{overlapped}}$, where $n_{overlapped}$ is the number of shared pixels.

The median pointing uncertainty (i.e. off$_{(i)}$ in eq~\ref{eq:cross}) is estimated to be $3.2''$ for individual maps after being corrected by pointing observations and before corrections presented here, and the median zero-flux uncertainty is 1 mJy/beam.

\subsection{Processing of \textit{Herschel}, \textit{Planck} and CSO/Bolocam maps}

\begin{figure*}
\center
\includegraphics[scale=0.23]{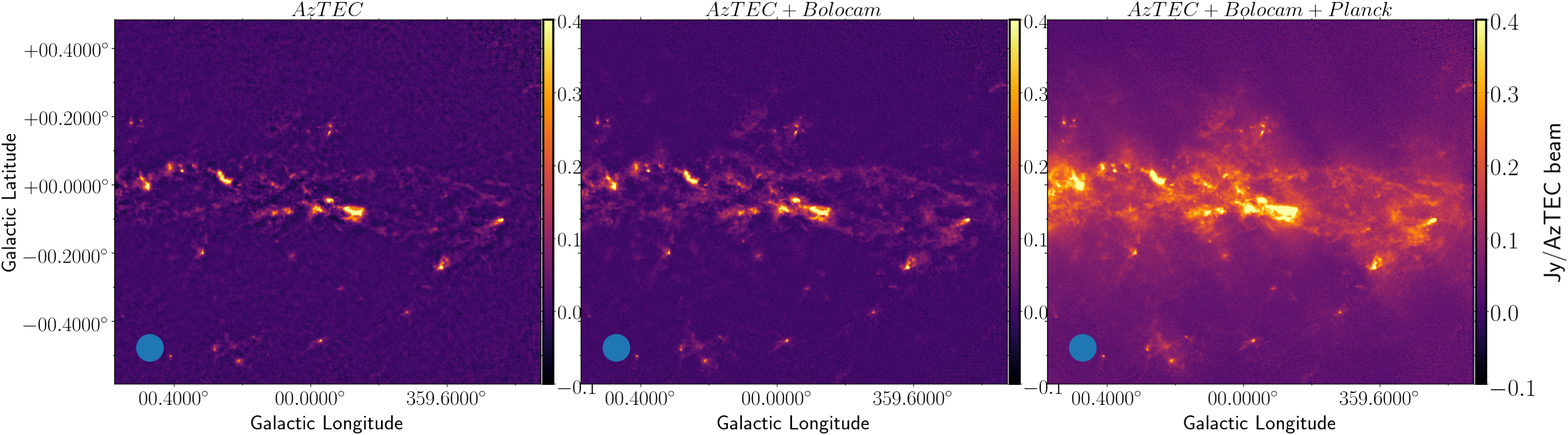}

\caption{Maps at different combining stages. Left: Same as Figure~\ref{fig:raw_map}, the AzTEC map; Middle: the AzTEC+Bolocam map; Right: the final combined map. The CO 3-2 contamination has been removed from the \textit{Planck} map. This procedure relies on precise knowledge of different transmission curves in individual bolometers and suffers from systematic noise, which manifests as stripe-like artefacts visible in the image. The \textit{Planck}/HFI beam is indicated in the bottom-left corner.}
\label{fig:map_comb}
\end{figure*}

\begin{figure}
\center
\includegraphics[scale=0.4]{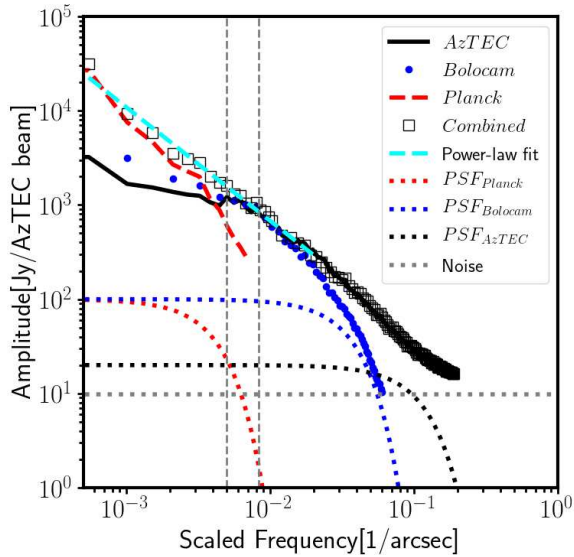}

\caption{The ASDs of individual maps at 1.1 $mm$ and that of the combined map. The units in all maps are converted to $Jy/(9.5'' beam)$. The \textit{Planck}/HFI 353 GHz map is shifted to $\lambda=1.1$ $mm$, using flux ratios between $F_{1.1mm}$ and $F_{850\mu m}$ derived from iterative fitting to $160-850$ $\mu m$ maps. The frequency has been scaled by a factor of $\frac{\pi}{2ln(2)}$, such that $f_{1/2} = \frac{1}{FWHM}$ for a Gaussian beam (See text for details). The spectral densities of all maps are truncated at a frequency corresponding to a spatial scale of $\approx 0.5 \times FWHM$. The green dotted line shows a power-law fit to the ASD of the combined map below $1/50''$. The horizontal line indicates the ASD of the white noise in the AzTEC map (15 mJy/beam). The two vertical lines indicate the sizes of the Bolocam's (left) and the AzTEC's (right) detector arrays.} 
\label{fig:asd_maps}
\end{figure}

\subsubsection{Map Combination}

We create a combined $1.1$ $mm$ map from the AzTEC $1.1$ $mm$ map, the \textit{Planck}/HFI $353$ GHz map (\textit{Planck} 2013 data release (PR1)) and the CSO/Bolocam $1.1$ $mm$ map \citep{aguirre11,ginsburg13}, to compensate for the large scale emission filtered out by the PCA cleaning in the AzTEC map. The \textit{Planck}/HFI ($353$ GHz or $850$ $\mu m$) map is scaled to 1.1 mm to match the wavelength of the AzTEC \& Bolocam maps before merging. The $850$ $\mu m$ intensities are extrapolated to $1.1$ $mm$ after applying a pixel-by-pixel SED fitting to the 160-850 $\mu m$ maps (See Section~\ref{sec:ccrr}). 
The \textit{Planck} $353$ GHz map is contaminated by CO J=3-2 emission, and so the estimated CO J=3-2 map was downloaded from the \textit{Planck} Legacy Archive and subtracted from the original map. Here we use their Type 1 correction map, which is based on the precise knowledge of different filter responses in each individual bolometer with no further assumption on line ratios. This estimated CO J=3-2 map is subtracted from the original map. The CO J=3-2 line emission could contribute up to $30\%$ of the total flux in the broad continuum filter. The net \textit{Planck} map after the CO contribution removal shows large-scale, artificial stripes introduced by this process (Figure~\ref{fig:map_comb}). 

The combined map is created using an approach described by \cite{faridani18}. This approach is mathematically equivalent to "feathering", an approach widely used for combining interferometer observations with single-dish observations, but is operated in the ``map domain'' instead of the Fourier domain. To combine the AzTEC map and the Bolocam map:

\begin{equation}
\mathbf{M(AB)} = \mathbf{M(A)} + [\mathbf{M(B)} - \mathbf{(M(A)} \otimes \mathbf{beam(B-A)})] 
\end{equation}

We degrade the resolution of the AzTEC map $\mathbf{M(A)}$ ($10.5''$) to that of the Bolocam map $\mathbf{M(B)}$ ($33''$), where $\otimes$ stands for convolution, and $ \mathbf{beam(B-A)}$ stands for a Gaussian kernel with $\sigma_{B-A} = \sqrt{(33^2 - 10.5^2)}''$. The peak fluxes of major clouds are compared between the degraded AzTEC map and the Bolocam map, after color corrections, which suggests a calibration offset of $10\%$. The AzTEC map is multiplied by a factor of 1.1 to match to the Bolocam map, since as mentioned in Section~\ref{subsec:mapping}, our AzTEC is likely under-calibrated due to a lack of flux-calibrators observed at the low elevations where the CMZ is observed. The degraded AzTEC map is then subtracted from the Bolocam map. This differential map with an ``excess'' component at the scale of the Bolocam beam ($33''$) is added back to the original AzTEC map to produce a 
combined map $\mathbf{M(AB)}$. 

Similarly, to further incorporate the \textit{Planck} map:

\begin{equation}
\mathbf{M(ABP)} = \mathbf{M(AB)} + [\mathbf{M(P)} - \mathbf{(M(AB)} \otimes \mathbf{beam(P-B)})] 
\end{equation}

The amplitude spectral densities (ASDs) of individual maps and that of the combined image are shown in Figure~\ref{fig:asd_maps}, where ``amplitude'' $|F(u,v)|$ is the azimuthally averaged, absolute value of the Fourier transform of each map:
\begin{equation}
F(u,v)=\sum_{k=0}^{n_k}\sum_{j=0}^{n_j}Map[k,j]exp\left[-2\pi i(\frac{u \times k}{n_k}+\frac{v \times j}{n_j})\right] 
\end{equation}
The frequency (i.e., $\sqrt{u^2+v^2}$) is multiplied by a factor of $\frac{\pi}{2ln(2)}$, in this way, a Gaussian beam drops to half of its peak value in the Fourier domain at a frequency of $f_{1/2}=\frac{1}{FWHM}$, where FWHM is defined in the ``map'' domain. The lowest recoverable spatial frequency in the AzTEC and that in the Bolocam maps are similar, $\approx \frac{1}{100''}$, as consequences of the PCA cleaning, which removes correlated signals across the entire array (i.e., signals with spatial scale larger than the size of the detector array). On the other hand, the highest recoverable spatial frequency in the \textit{Planck} map is limited by its FWHM $=292''$ PSF. Therefore, a gap emerges between $1/300''-1/100''$. Nevertheless, the ASD of the combined map suggests that the spectrum approximately follows a power law. A power-law fit to the ASD of the combined map at frequencies below $1/50''$ yields an exponent of $-1.2$ and a standard deviation of $12\%$ (cyan dashed line in Figure~\ref{fig:asd_maps}).

The $1\sigma$ statistical noise is $15$ mJy/beam in the AzTEC map. In terms of the surface brightness, the random noise in the Bolocam map and that in the \textit{Planck}/HFI map is negligible. We further apply a $10\%$ relative calibration uncertainty to the final 1.1 $mm$ combined map, which accounts for beam variations in the AzTEC maps. 

The errors in the \textit{Herschel} maps are dominated by calibration uncertainties, and can be divided into relative calibration uncertainties and absolute calibration uncertainties. We adopt a relative calibration uncertainty of $2\%$ for all SPIRE bands, and $5\%$ for the PACS $160$ $\mu m$ band \citep{bendo13, balog14}. We notice that some authors adopted more conservative estimates of the relative uncertainties for extended sources, inferred from comparisons between observations taken by \textit{Herschel}/PACS and those made by other facilities. (e.g. Spitzer/MIPS, AKARI, \citep{juvela15}). However, the fluctuations in the low surface brightness region of the PACS $160\mu m$ maps indicate that the relative uncertainty should be $<5\%$. Furthermore, it is dangerous to model absolute calibration offsets without an accurate knowledge of the dust absorption curve since these two ingredients are completely degenerate. When absolute calibration offsets are modeled as free parameters, we obtain a best-fit offset of $20\%$ at 160 $\mu m$, and meanwhile systematically higher dust spectral indices (by 0.3) compared to those derived from a data set with calibration offsets fixed to zero. This large calibration offset could be alternatively a misrepresentation of our simplified dust model, which has a single temperature and a single power-law absorption curve. Without a precise knowledge of the model uncertainty, we decide to fix absolution calibration offsets to zero instead of trying to discern between the calibration uncertainty and the model uncertainty, assuming that the former is better understood.

\subsubsection{Color Corrections and Flux Ratios $F_{1.1mm}$/$F_{850\mu m}$}\label{sec:ccrr}

\begin{figure*}
\center
\includegraphics[scale=0.28]{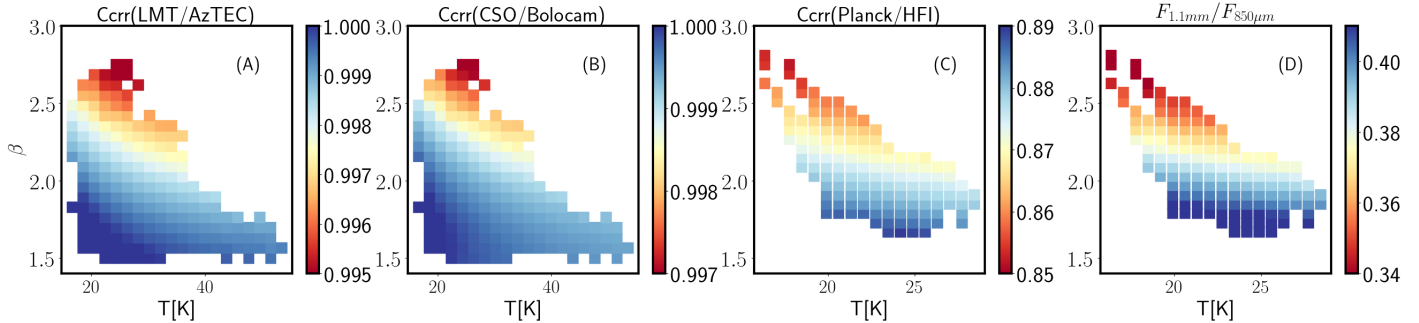}

\caption{(A)-(C):Color corrections for the AzTEC, the Bolocam and the \textit{Planck}/HFI maps. Plotted are all parameter bins containing at least one cell/pixel. The correction factors are calculated from eq~\ref{eq:ccrr} and best-fit modified blackbody SEDs of uncorrected $160$ $\mu m$-1.1 mm maps (AzTEC/Bolocam) or $160-850$ $\mu m$ maps (\textit{Planck}/HFI). The small relative variations across all maps suggest that no further iteration is needed. (D): Flux ratios between $F_{1.1mm}$ and $F_{850\mu m}$, which are extimated from best-fit SEDs of color-corrected 160-850 $\mu m$ maps.}

\label{fig:ccrr}
\end{figure*}

The color correction factor for a source with a spectrum $F_{src, \nu}$, when observed with a broad-band photometer characterized by a system spectral bandpass $t_{\nu}$, is defined as:

\begin{equation}
Ccrr = \frac{F_{cal, \nu_c}}{\int_{\nu}F_{cal, \nu}t_{\nu}d\nu} \left[ \frac{F_{src, \nu_c}}{\int_{\nu}F_{src, \nu}t_{\nu}d\nu} \right]^{-1}
\end{equation}\label{eq:ccrr}

\noindent where $F_{cal, \nu}$ is the spectrum assumed during the calibration process to convert from the integrated flux over the broad-band to the monochromatic flux density at band center $\nu_c$ (e.g., $\nu_c$ = 1.1 mm for the AzTEC and the Bolocam). The \textit{Planck} and the \textit{Herschel} adopt a non-thermal, ``flat'' spectrum: $F_{cal,\nu}\nu = const$ or $F_{cal, \nu} \propto \nu^{-1}$. This results in non-trivial color corrections for sources with a thermal SED (e.g., $Ccrr \lesssim 95\%$ for \textit{Herschel}/SPIRE\footnote{\url{http://herschel.esac.esa.int/Docs/SPIRE/spire_handbook.pdf}} at 500 $\mu m$ and \textit{Planck}/HFI\footnote{\url{https://wiki.cosmos.esa.int/planckpla2015/index.php/Unit_conversion_and_Color_correction}} at 1.1 mm). In contrast, both AzTEC and Bolocam~\citep{aguirre11} are calibrated with thermal/modified blackbody spectral, and have $Ccrr \approx 1$.

Figure~\ref{fig:ccrr} shows the color corrections for the AzTEC, the Bolocam and the \textit{Planck}/HFI maps, which are determined from eq~\ref{eq:ccrr} and best-fit dust SEDs (see Section~\ref{sec:deconvol}) of the uncorrected 160 $\mu m$-1.1 mm combined maps (for AzTEC/Bolocam) image or the 160-850 $\mu m$ maps (for \textit{Planck}/HFI). The multi-band maps are first fitted with a modified black-body model without color correction to obtain an estimate of $F_{src,\nu}$ in eq~\ref{eq:ccrr}. The system spectral bandpass for the Bolocam is obtained from~\cite{aguirre11}. For the \textit{Planck}/HFI data, we take use of the \textit{Planck} unit conversion and colour correction software\footnote{\url{http://pla.esac.esa.int/pla}}. The above procedure also provides an estimate of flux ratios $F_{1.1mm}$/$F_{850\mu m}$ for each pixel to convert from the \textit{Planck}/HFI map to 1.1 mm. For both color corrections and $F_{1.1mm}$/$F_{850\mu m}$, the relative variations across the map are small $1\sigma < 3\%$, suggesting that no further iteration process is needed.    

To construct dust SEDs, we also take advantage of existing \textit{Herschel} PACS/SPIRE $160$ $\mu m$, $250$ $\mu m$, $350$ $\mu m$ and $500$ $\mu m$ maps from the Hi-GAL survey \citep{molinari10}, we downloaded the Photometer Calibration Products\footnote{\url{http://herschel.esac.esa.int/twiki/bin/view/Public/SpireCalibrationProducts}} from the ESA \textit{Herschel} Science Archive for color corrections, which contains filter responses and aperture efficiencies for each band. For PACS 160 $\mu m$, the color correction factor for each pixel is then calculated again by eq~\ref{eq:ccrr}. For SPIRE maps, which are calibrated in Jy/beam, a further convolution of the position-position-frequency SED cube with a frequency-dependent beam profile is needed. This approach is crucial for compact sources with sizes comparable to the sizes of the \textit{Herschel}/SPIRE PSFs (e.g., Sgr C), which cannot be approximated as either a single point source or a smooth beam-filling source. We verify that for given spectral shapes (parameterized by temperature T and spectral index $\beta$), our inferred color corrections are consistent with tabulated values in the PACS and SPIRE handbooks.

\section{Analysis: Model-based Deconvolution with MCMC} \label{sec:deconvol}

Throughout this project, we use Bayesian analysis on the dust SEDs to explore physical properties of dust grains. A common problem in SED analysis is non-uniform resolution across multi-band images. The conventional approach to address this issue is to convolve every image to the lowest resolution (C2LR, hereafter). In this work, we adopt a forward modeling strategy to fit a dust model to multi-band maps, each diluted by a different instrumental PSF (i.e., a model-based deconvolution, MBD hereafter). The goal is to achieve optimal spatial resolution for each parameter being fit. To demonstrate this approach, we start with a simple model, assuming that the dust SED follows a standard modified blackbody form with a single temperature approximation. The wavelength-dependent dust absorption curve $\kappa_{\lambda}$  is assumed to be a single power-law, characterized by a spectral index $\beta$. This model is referred to as STMB (single temperature modified blackbody) hereafter. 

\subsection{Dust Model}
A cellular-based STMB model for the entire CMZ relies on three parameter grids: a temperature grid $\boldsymbol{T}$,  a column density grid $\boldsymbol{N_{H_2}}$ and a grid of the dust spectral index $\boldsymbol{\beta}$. The surface brightness $F_{i}(\nu_j)$ at pixel, i, and frequency $\nu_j$ is given by:

\begin{equation}
F_{i}(\nu_j) = [1-exp(-\tau_{i,\nu_j})] B_{\nu_j}(T_{i}) \Omega_j
\end{equation}
\label{eq:modbk}

\noindent where $\Omega_j$ is the beam area in the $j$th band. 
$B_{\nu_j}(T_{i})$ is the \textit{Planck} function. ${\tau}_{i,\nu_j}$ is the optical depth at frequency $\nu_j$, which is given by:

\begin{equation}
\tau_{i,\nu_j} = \kappa_0 (\frac{\nu_j}{\nu_0})^{\beta_{i}} \mu m_{H} \times {N_{H_2}}_{i} \times 1\%   
\end{equation}
\label{eq:tau}

\noindent where $\kappa_0$ is the absorption cross section per 
unit mass at frequency $\nu_0$. We adopt $\kappa_0=1.37$ cm$^2/g$ and $\nu_0=c/1000$ $\mu m$ from \cite{ossenkopf94} for coagulated dust grains with thin ice mantles (their Table 1), $c$ is the speed of light. We also adopt a mean molecular weight $\mu=2.8$ and a dust-to-gas mass ratio of $1\%$ to convert from $N_{H_2}$ to column dust mass density. This model is not restricted to an optically thin approximation ($\kappa \propto \nu^{\beta}$).

The calculated raw flux map $\mathbf{F(\nu_j)}$ is diluted to the instrumental resolution of each wavelength band to match the data:
\begin{equation}
\mathbf{Model(\nu_j)} = \mathbf{F(\nu_j)} \otimes \mathbf{beam_{j}}
\end{equation}

\noindent where $\otimes$ refers to convolution. All beam profiles are approximated as Gaussian profiles. The FWHM of the beams are $13.6''$ at $160\mu m$, $23.4''$ at $250$ $\mu m$, $30.3''$ at $350$ $\mu m$, $42.5''$ at $500$ $\mu m$ and $10.5''$ at $1.1$ $mm$, respectively. Note that the beam sizes of the PACS/SPIRE maps are larger than their nominal values \citep{traficante11}, which is due to the high scanning speed ($60''/s$) adopted by the Hi-Gal survey.

\subsection{Sampling Strategy} \label{sec:MCMC}

\begin{figure}
\center
\includegraphics[scale=0.5]{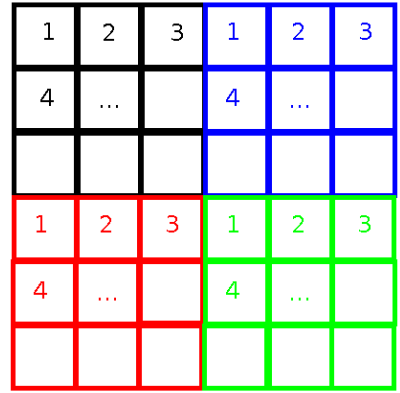}

\caption{An illustration of how sampling procedure is parallelized. The $6 \times 6$ grid is divided into four $3 \times 3$ sub-blocks. At step 1, all cells with index 1 are updated in parallel, at step 2, all cells with index 2 are updated in parallel. In this case, this procedure guarantees that any 2 cells updated simultaneously have at least a separation of 3 cells.}
\label{fig:parall}
\end{figure}

We perform Markov chain Monte Carlo (MCMC) analysis to derive optimal parameters for our model, using a Slice-within-Gibbs sampling strategy\citep{neal03}. A step-by-step recipe is provided in Appendix~\ref{apendix:slice}. The advantage of the Gibbs sampling strategy is that, since one single parameter (of a certain grid cell) is updated per step, we can avoid frequent convolutions of each entire image with its PSF, which is computationally expensive. Instead, the global likelihood is modified within an area of $8$ $\sigma_{PSF}$ around each parameter cell to be sampled. The sampling procedure is parallelized by dividing each parameter grid into multiple identical sub-blocks. The block size must be larger than $8\sigma_{PSF, max}$, where $\sigma_{PSF, max}$ is $\sigma$ of the largest beam. Updating is looped over all sub-blocks in parallel, as illustrated by Figure~\ref{fig:parall}, which guarantees that simultaneously updated cells are always separated by $d>8\sigma_{PSF, max}$. 

A new updating is always accepted by a Gibbs sampler, while a multivariate random-walk Metropolis updating at such high dimensionality could be extremely inefficient due to its low acceptance rate. We have developed a C++ package to perform this MCMC analysis. The source code of this package will be made public along with a future paper focusing on densities and temperature structures of clouds in the CMZ. The sampler is run for 4000 steps for every parameter. The convergence is examined by both trace plots and auto-correlations, and is rapidly achieved after a few hundred steps. The optimal value of each parameter is derived by the median value of the last 2000 steps. The computation time depends on the size of the image/parameter grid and the sizes of the PSFs. For a image/grid size of $640 \times 320$ pixel/cell and a maximum PSF size of $21 \times 21$ pixel at 500 $\mu m$, our analysis takes 22 hrs for 4000 steps on a workstation with 25 Intel Xeon E5-2660 2.20 GHz CPU threads (i.e., parallelized over $5 \times 5$ sub-blocks).

Our procedure shares the same advantage of the PPMAP procedure applied to previous Hi-GAL surveys \citep{marsh15},  meaning that spatial resolutions of the best-fit parameter grids are improved by incorporating the PSF knowledge into the model. PPMAP assumes an optically thin condition in order to retain the linearity and solvability of the model. By using a Gibbs sampler, we can avoid frequent convolutions of the entire map, and an optically thin condition is not mandatory. On the other hand, we do not have a large number of independent temperature components along the line of sight, which can be achieved in the PPMAP.

\subsection{Smoothness Prior} \label{sec:prior}

Depending on the ratios between the cellular size of the parameter grid and the sizes of the PSFs, overfitting to noise can occur, which manifests as high-frequency fluctuations among neighboring cells. For $T$ and $\beta$, this issue is more serious, as both parameters are constrained by multiple bands. In those bands (e.g., 500 $\mu m$) where the PSF is over-sampled, the effective degrees of freedom are reduced due to increased number of free-parameters in contrast to a lack of extra information on sub-PSF scales. Conversely, $N_{H_2}$ suffers less from overfitting since it is predominantly determined by the single highest-resolution map at 1.1 $mm$. Overfitting could be mitigated by regularized Bayesian inference ~\citep{warren03}.  Here, we choose the simplest gradient form of regularization, which can be expressed as:

\begin{equation} \label{eq:smprior}
ln(P(x_{ix,iy})) = ln(L(x_{ix,iy})) - \lambda G \\
\end{equation}

\begin{equation} 
G =  \sum\limits_{+,-} \frac{(x_{ix,iy} - x_{ix\pm1,iy})^2 +  (x_{ix,iy} - x_{ix,iy\pm1})^2}{2}
\end{equation}

\noindent $ln(P(x_{ix,iy}))$ and $ln(L(x_{ix,iy}))$ are the logarithms of the full conditional posterior and likelihood for parameter $x_{ix,iy}$, where $x$ is $\{N_{H_2},T,\beta \}$. The regularization term G can be regarded as an \textit{a priori} smoothness of parameter $x$. $\lambda$ is a user-defined weight of the smoothness prior. One can rewrite $\lambda$ as $\lambda = \frac{1}{2\sigma^2}$, where $\sigma$ can be viewed as an \textit{a priori} mean standard deviation of adjacent cell-cell differences.

Our choice of $\sigma$ or $\lambda$ is guided by performing this regularized Bayesian analysis on simulated multi-band images of a cloud with known $N_{H_2}$, $T$ and $\beta$ distributions resembling that of the realistic CMZ cloud. We take advantage of \textit{Herschel} observations toward a nearby massive molecular cloud, M17. This cloud is selected by its high mass ($>10^5 M_{\odot}$, \cite{rice16}) and close distance (2.1 kpc, \cite{povich09}). At this distance, M17 is resolved down to a physical scale of $\approx0.4$ pc at $500$ $\mu m$ (FWHM = $42.5''$), which is roughly equal to that of a CMZ cloud at 1.1 mm (FWHM = $10.5''$). Below, we briefly describe how we utilize observations toward M17 to explore optimal choices of $\sigma$ (or $\lambda$ in Eq~\ref{eq:smprior}) for a CMZ-like cloud. For more details, see Appendix~\ref{apendix:prior}.

\begin{figure*}
\begin{center}
\includegraphics[scale=0.4]{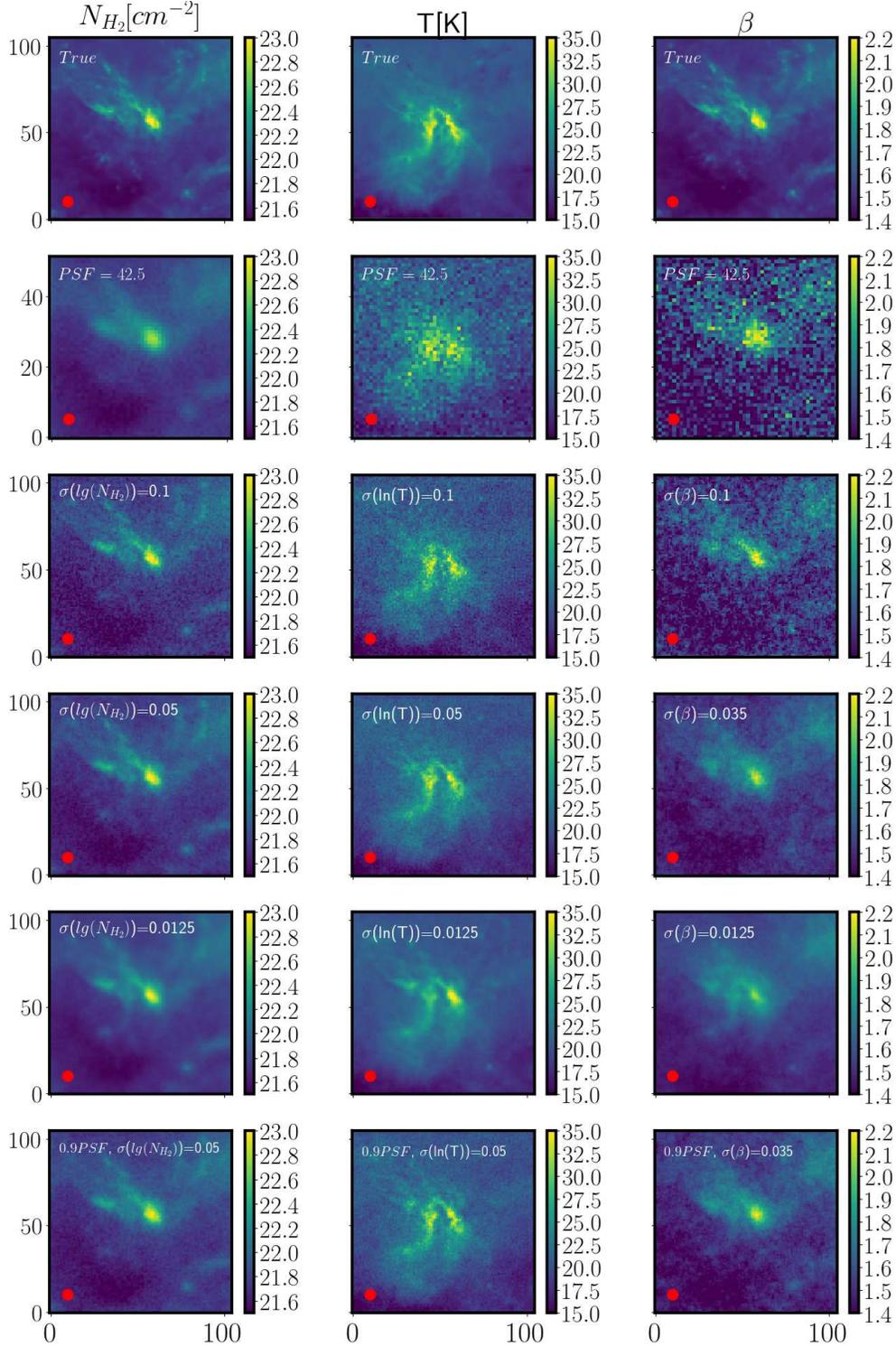}
\end{center}
\caption{The ``true'' maps of $lg(N_{H_2})$, T, and $\beta$ (top row),together with various ``best-fit'' maps derived via different approaches (subsequent rows; one for each). The SEDs are simulated from the ``true'' maps of $lg(N_{H_2})$, $T$, assuming that $\beta$ as a function of $N_{H_2}$ (see text for details) and that the cloud was at a distance of 8.4 kpc. The ``true'' maps of $lg(N_{H_2})$ and $T$ are inferred from \textit{Herschel} $160-500$ $\mu m$ maps of M17 at its original distance, 2.1 kpc, but rebinned by a factor of 4. $\beta$ is assumed to follow a broken power-law as a function of $N_{H_2}$. The second-row shows results derived by degrading every map to $FWHM=42.5''$ (the C2LR approach). The third to fifth rows show results derived with the MBD approach, using different smoothness priors. The sixth-row shows results derived with incorrect PSFs, assuming that the PSFs in all bands from 160 $\mu m$ to 1.1 $mm$ are underestimated by $10\%$. The maps derived with the C2LR approach (the second row) have a dimension of $52 \times 52$ and a pixel scale of $14''$. Other maps have a dimension of $105 \times 105$ and a pixel scale of $7''$. The largest beam ($FWHM=42.5$) is shown in the lower-left corner.}
\label{fig:imgs_train}
\end{figure*}

We start by constructing ``true'' maps of $N_{H_2}$, $T$, and $\beta$ based on \textit{Herschel} images of M17, from which we simulate $160$ $\mu m$ to 1.1 mm images of M17, assuming that the cloud is at the distance of the CMZ ($\approx8.4$ kpc). After this, we perform our regularized Bayesian analysis on the simulated multi-band images with different strengths of smoothness prior. The best-fit maps of $lg(N_{H_2})$\footnote{$lg$ means base-10 $log$}, $T$ and $\beta$ under different priors are plotted in Figure~\ref{fig:imgs_train}. Different priors can be compared more quantitatively by inspection of the amplitude spectral densities of their corresponding best-fit images. As shown in Appendix~\ref{apendix:prior}, for a grid cell size of $7''$, a combination of $\sigma(lg(N_{H_2[cm^{-2}]}))=0.05$, $\sigma(ln(T[K]))=0.05$ and $\sigma(\beta)=0.035$ achieves a reasonable compromise between noise reduction and preservation of resolution. We adopt this smoothness prior throughout all of the following analyses. 

\section{Results} \label{sec:results}

\subsection{High-Resolution Maps of $N_{H_2}$, $T$ and $\beta$}

\begin{figure*}
\begin{center}
\includegraphics[scale=0.45]{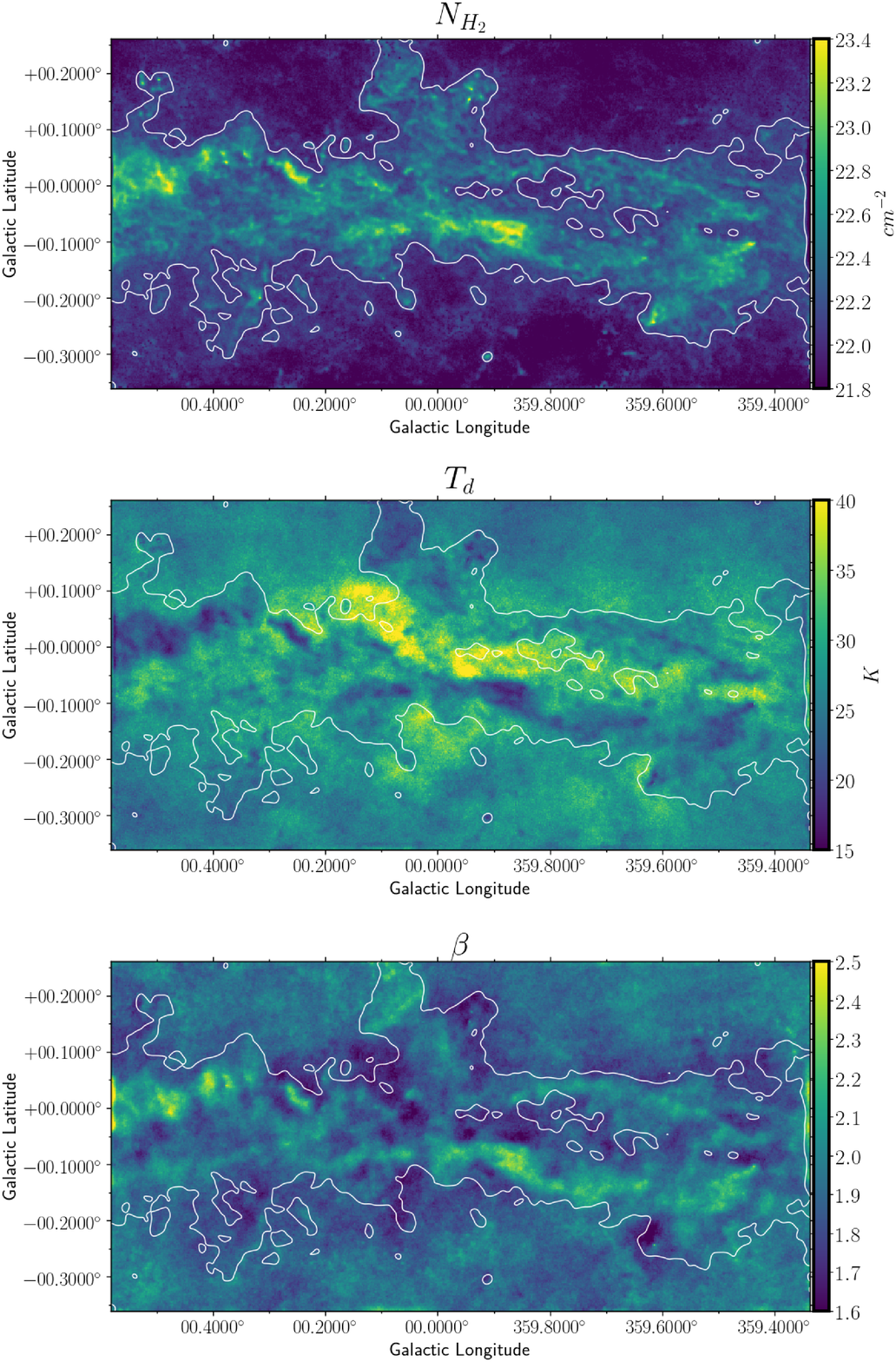}
\end{center}
\caption{Best-fitted maps of $N_{H_2}$, $T$ and $\beta$ for the CMZ. All maps have a grid cell size of $7''$. The white contour indicates a level of $N_{H_2}=10^{22.1}cm^{-2}$, which corresponds $\approx 6\sigma$ at 1.1 mm. In this analysis, we adopt a smoothness prior of $\sigma(lg(N_{H_2[cm^{-2}]}))=0.05$, $\sigma(ln(T[K]))=0.05$ and $\sigma(\beta)=0.035$. Readers should be cautious about the structures at low column densities, where flux densities at 1.1 mm are solely recovered from \textit{Planck} observations and have much greater PSFs ($292''$) than assumed ($10.5''$).}
\label{fig:maps_fit}
\end{figure*}

\begin{figure*}
\begin{center}
\includegraphics[scale=0.35]{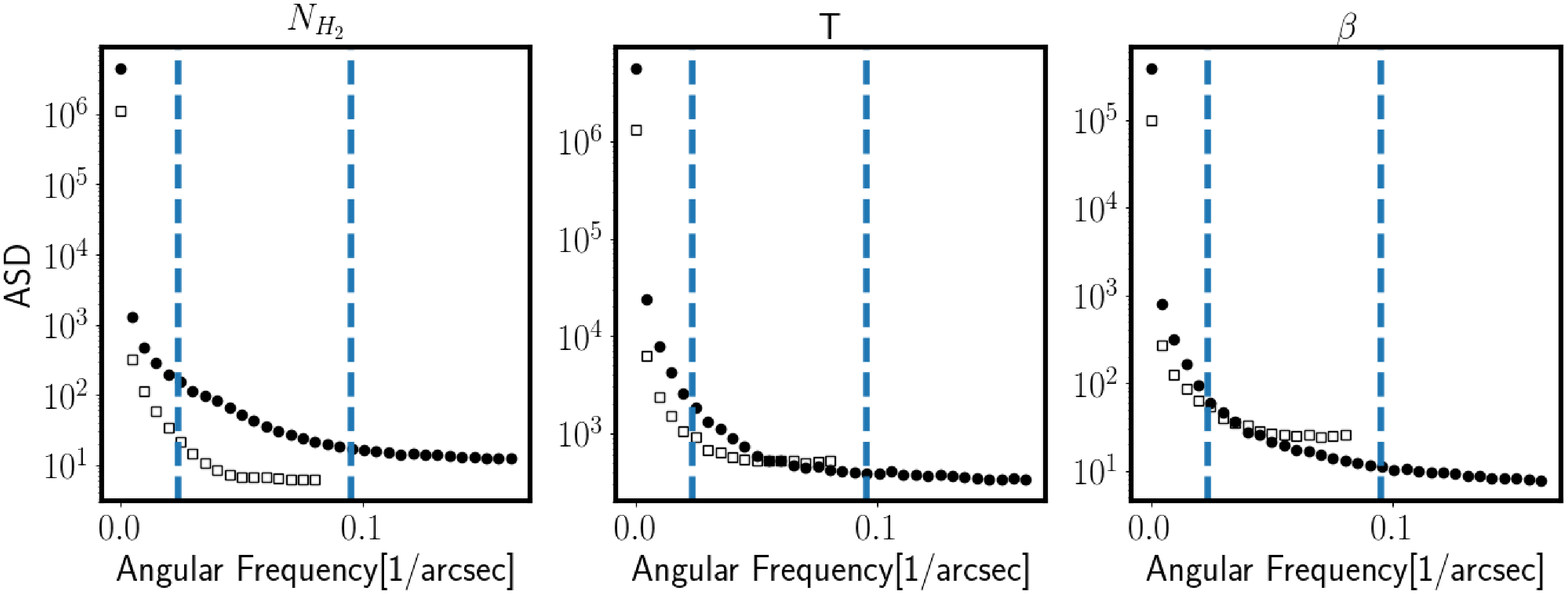}
\caption{The ASDs of each individual image shown in Figure~\ref{fig:maps_fit}. The ASDs derived by degrading the individual maps to FWHM$=42.5''$ are also shown in squares for comparison. In each plot, the vertical line on the left indicates the largest PSF ($FWHM=42.5''$ at $500\mu m$), and the vertical line on the right indicates the smallest PSF ($FWHM=10.5''$ at 1.1 mm). Notice that for the C2LR approach, parameter grids have a grid cell size of $14''$ instead of $7''$, which is still smaller than the Nyquist frequency of the $42.5''$ PSF.}
\end{center}
\label{fig:asd_fit}
\end{figure*}

\begin{figure*}
\begin{center}
\includegraphics[scale=0.6]{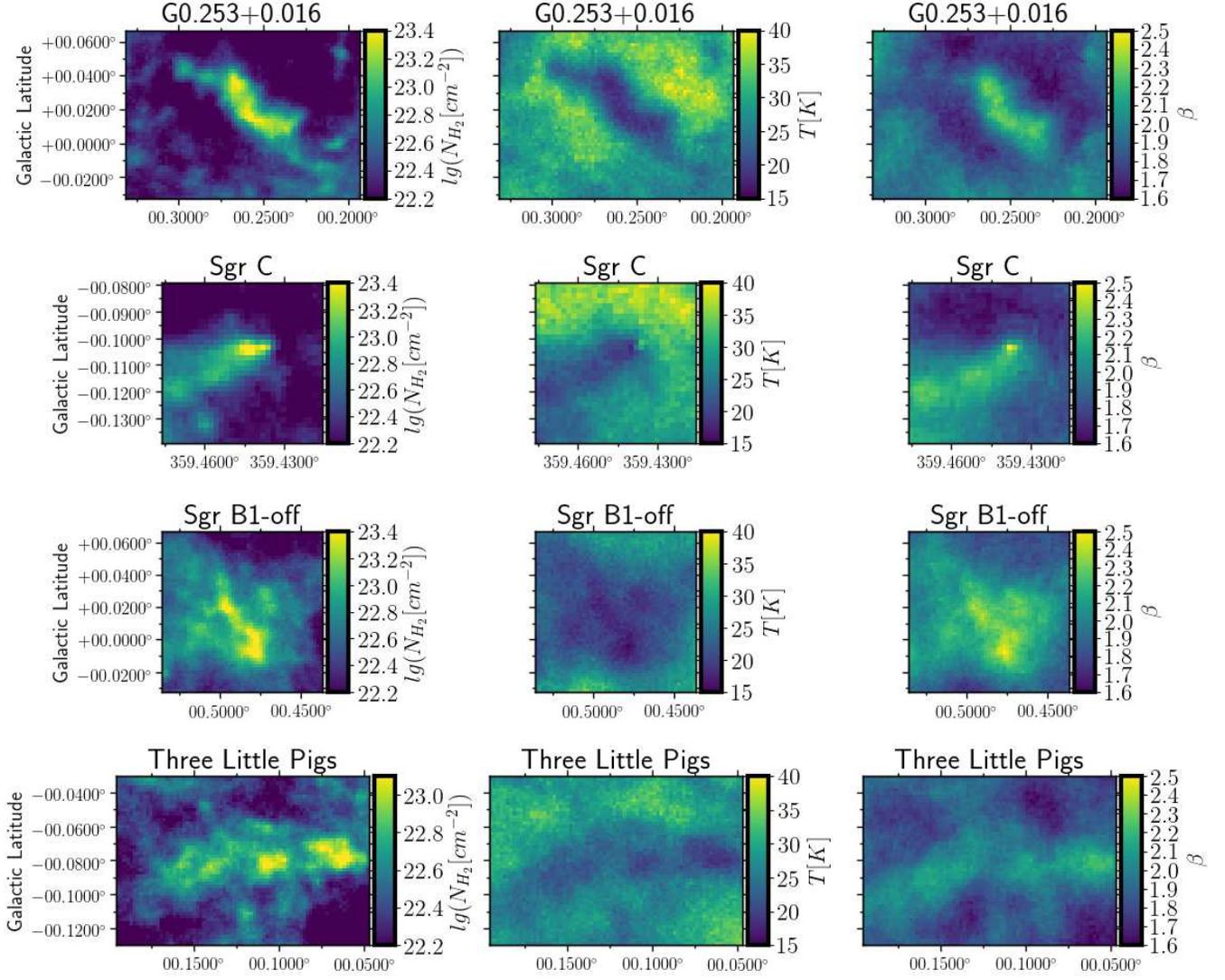}
\end{center}
\caption{Close-ups of four individual clouds/complexes: G0.253+0.016, Sgr C, Sgr B1-off, Three-Little-Pigs. Correlations between $N_{H_2}$, $T$ and $\beta$ are clearly shown. Studies of these correlations are present in a separate work.}
\label{fig:maps_sub}
\end{figure*}

\begin{figure*}
\begin{center}
\includegraphics[scale=0.8]{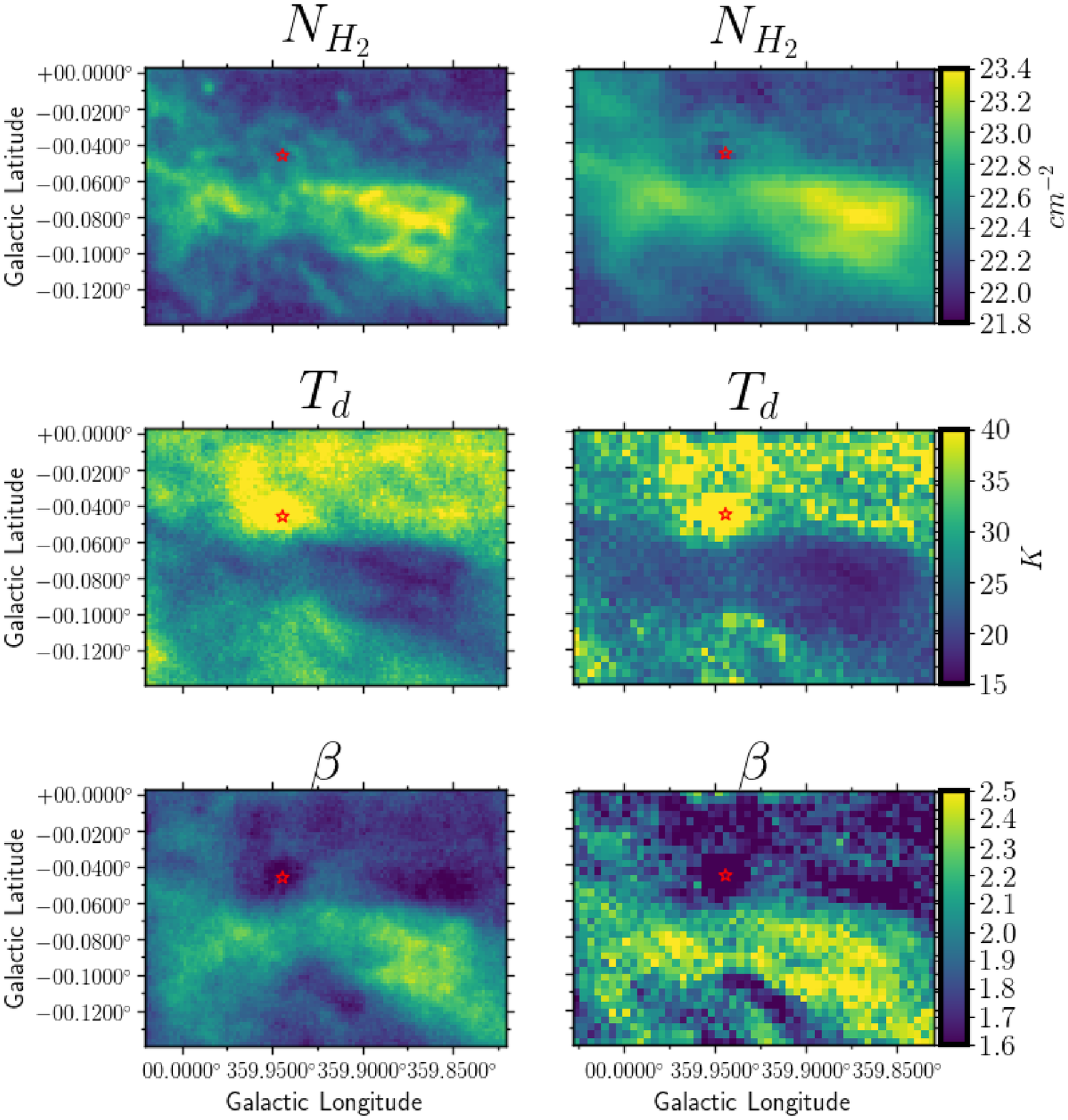}
\end{center}
\caption{Left: A close-up view of Figure~\ref{fig:maps_fit} on the 50 km/s and 20 km/s clouds. Right: Best-fitted maps of $N_{H_2}$, $T$ and $\beta$, derived with the C2LR approach. These maps have a pixel scale of $14''$ instead of $7''$. The red star show the position of Sgr A* for reference. Since Sgr A* is not a thermal source, its best-fit $N_{H_2}$, $T$ and $\beta$ are of less physical significance.}
\label{fig:maps_fitsub}
\end{figure*}

Figure~\ref{fig:maps_fit} shows best-fit maps of $N_{H_2}$, $T$ and $\beta$ for the CMZ, using the MBD approach described above. White contours indicate a column density level of $N_{H_2}=10^{22.1}$ cm$^{-2}$, which roughly corresponds to the $\approx 6\sigma$ detection limit of the AzTEC map as shown in Figure~\ref{fig:raw_map}. At low column densities, the 1.1 mm flux densities are solely determined by the CSO/Bolocam and \textit{Planck}/HFI maps, which have efficient resolutions lower than our assumed constant FWHM$=10.5''$ at 1.1 mm, potentially resulting in intermediate to large scale structures seen in $T$ and $\beta$ map at high latitudes. On the other hand, for the high-density CMZ region, our MBD approach, combined with the high-resolution AzTEC map, provides the finest large scale maps of $N_{H_2}$, $T$ and $\beta$ so far. 

The ASDs of the three maps are shown in Figure~\ref{fig:asd_fit}. As discussed in Appendix~\ref{apendix:prior}, to first order, the choice of $\sigma$ is justified from two considerations: a) the flattening of the ASDs due to noise fitting is seen only below the smallest PSF scale; and b) the ASDs derived with the MBD approach are higher than the ASDs derived with the C2LR approach on scales above the largest PSF scale. Both conditions are satisfied in Figure~\ref{fig:maps_fit}. 

The spatial resolution of our fitted parameter maps allows for studies of temperature and density structures close to clump scale ($\sim 1$pc), which are presented in a companion paper. Globally, there is a positive correlation between $\beta$ and $N_{H_2}$ and a negative correlation between $\beta$ and $T$. Interpretation of this trend requires proper foreground and background subtraction and careful understanding of correlated uncertainties (posteriors) between the fitted parameters. Also, we find that CMZ clouds generally have high $\beta$ values, with a maximum of $2.4$ towards dense peaks. We explore these features and the resulting implication on dust properties using hierarchical Bayesian analyses. 

Figure~\ref{fig:maps_sub} provides close-ups of four clouds/complexes in the CMZ and Figure~\ref{fig:maps_fitsub} provides a close-up comparison between the MBD approach and the C2LR approach on the 50 km/s (left) and the 20 km/s (right) clouds, which are two CMZ clouds close to SgrA* named by their local standard of rest (LSR) radial velocities. The improved spatial resolution of the former method is most apparent for $N_{H_2}$, which reveals a high density ``spine'' in the 20 km/s cloud. Radio emission from Sgr A*, between the two clouds, is misinterpreted by our model as a high temperature source. A future study of the CMZ with the TolTEC camera \citep{bryan18} with bands at 2.0 and 1.4 mm will possibly allow us to study free-free emission in the CMZ. Notice that with the C2LR approach, the grid cell size is set to $14''$. This is still lower than the smallest recoverable scale ($21.2''$) for the largest $42.5''$ PSF, which introduces an extra level of white noise in the best-fit ASDs below $21.2''$. 

\subsection{Monitoring of Sgr A*} \label{sec:res_monitor}

\begin{figure*}
\begin{center}
\includegraphics[scale=0.5]{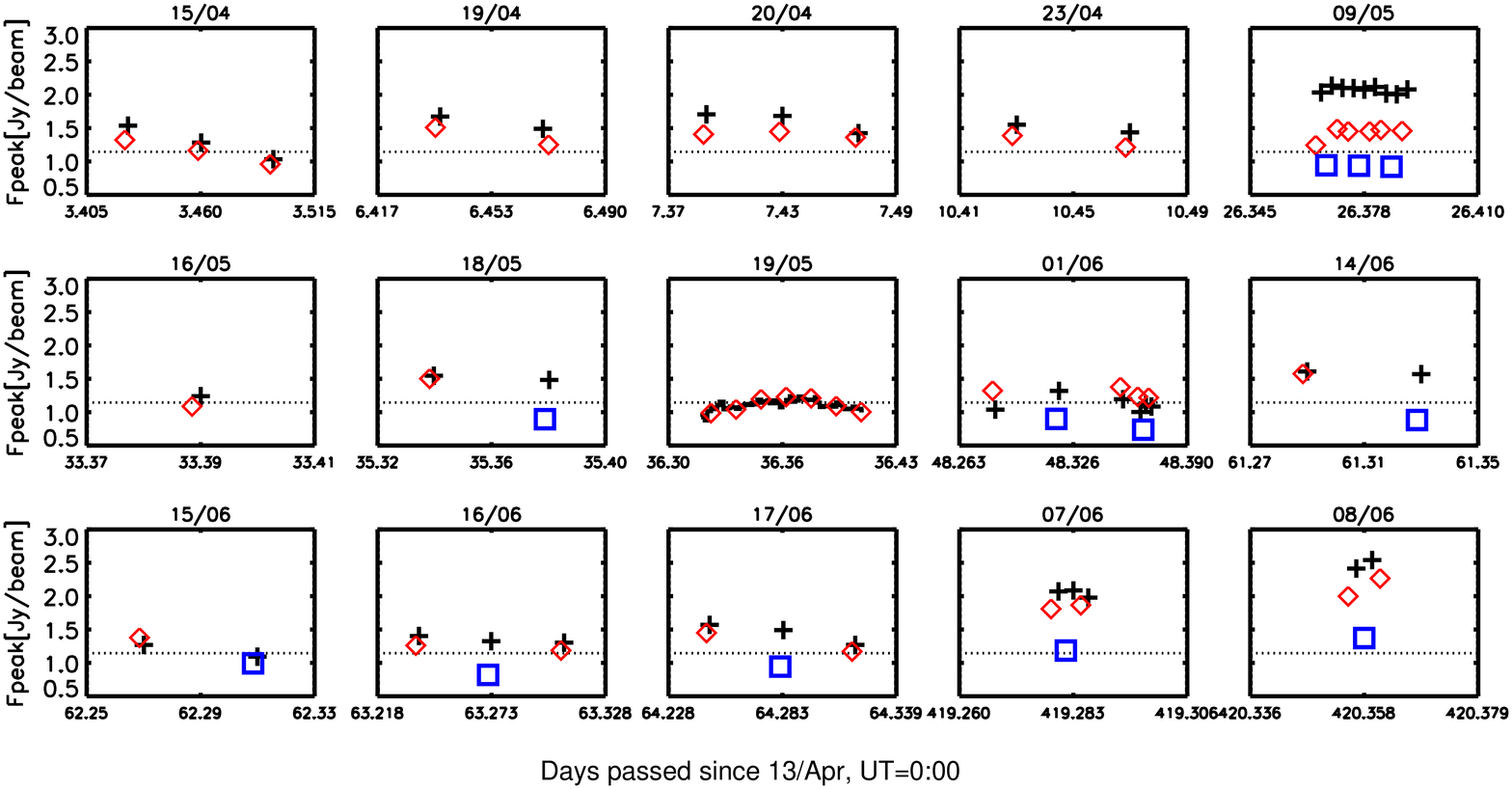}
\includegraphics[scale=0.5]{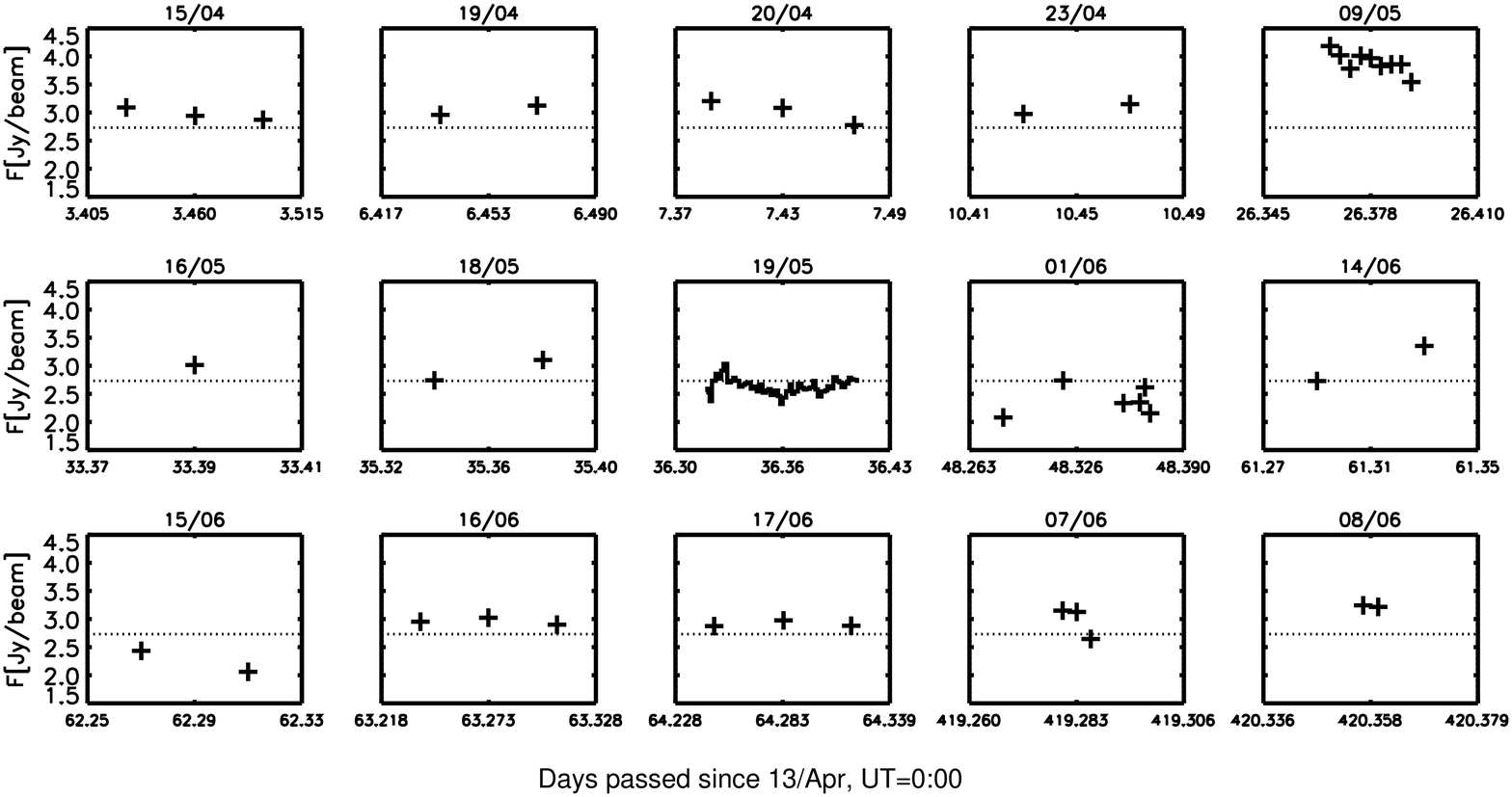}
\end{center}
\caption{Upper: The light curves of Sgr A* and calibrators during the 2014-2015 monitoring campaign. The black crosses indicate Sgr A*. The dense monitoring of Sgr A* on 19th May 2014 is shown by a solid line. The red diamonds and the blue squares indicate the two calibrators, BGPS G$000.378+00.041$ and AGAL G$359.984-00.009$, respectively. Each data point represents a 2 min integration. The fluxes of Sgr A* are divided by 2 for ease of comparison, and the fluxes of AGAL G$359.984-00.009$ are multiplied by 3. Observing dates are labeled on the top of individual panels. All sources are slightly smoothed with a Gaussian kernel of $FWHM=3''$ before $F_{peak}$ is measured. The absolute flux $F_{peak}$ is calibrated using a gain curve as a function of elevation and azimuth, which is determined from a series of observations towards asteroids throughout each season. Lower: The peak flux of Sgr A* is further calibrated relatively by the two calibrators shown in Figure 1. The calibration correction factor (CCF) is defined as: $Corr_{cal,i,j}=F_{cal,i,j}/median(F_{cal,j})$, where $F_{cal,i,j}$ is the $i$th observation of calibrator j. Each data point of Sgr A* is then multiplied by a linearly-interpolated CCF over observing time. The standard deviation of $F_{Sgr A*}$ is reduced from $20\%$ to $10\%$ after this relative calibration. An outburst is clearly detected on 9th May 2014. In both panels, the dotted line shows the median flux of Sgr A*, excluding the data points from 9th May 2014.}
\label{fig:lcurve_all}
\end{figure*}

\begin{figure}
\includegraphics[scale=0.35]{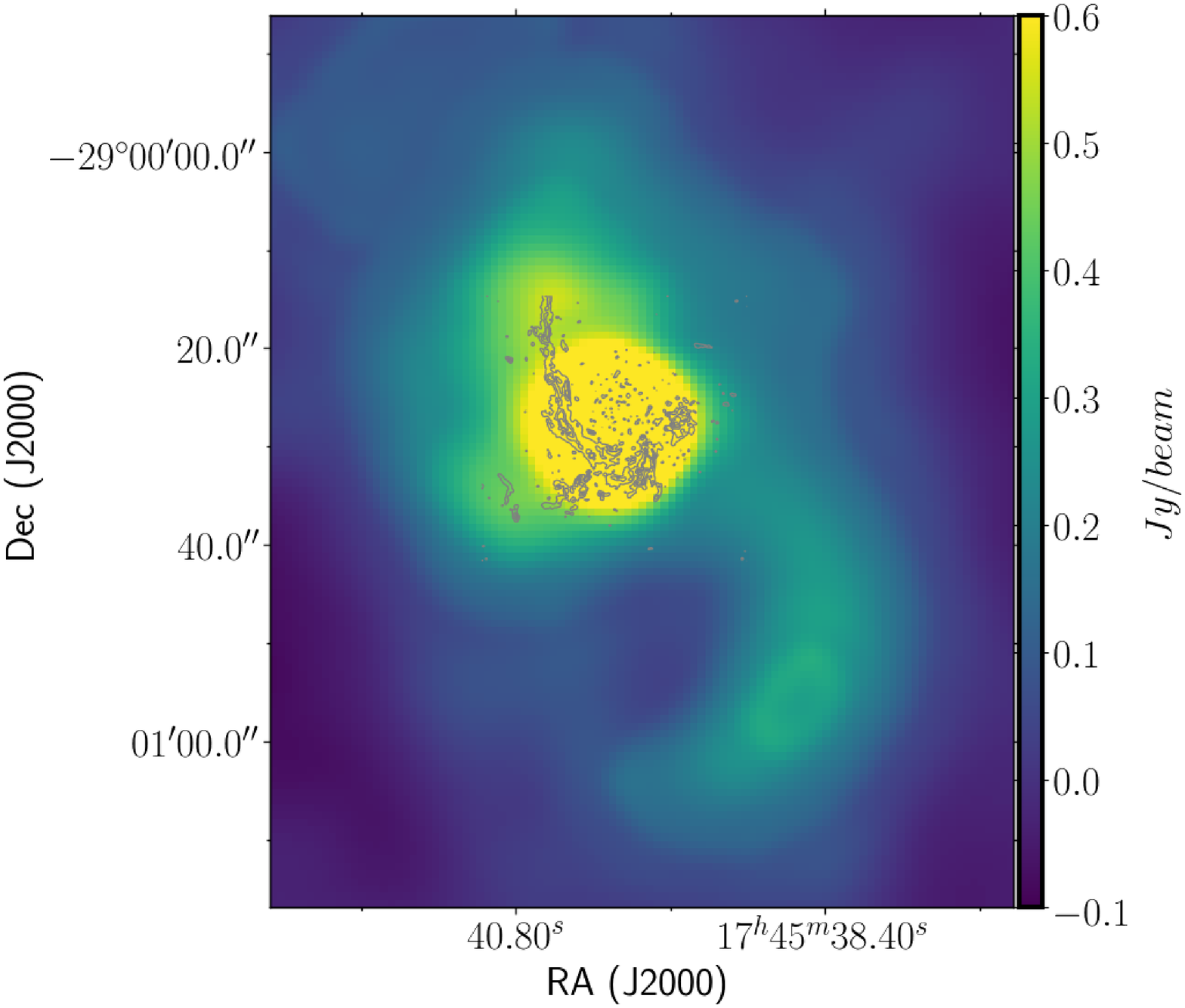}
\caption{LMT/AzTEC map of SgrA* with a $8.7''$ beam, constructed from a total of $\approx 7$ hour exposure taken from April 16 to June 17, 2014. The apparent 1.1 mm enhancements (NE and SE plumes) are seen in images generated with our data taken
on different days or with somewhat different methods. No significant variation of the enhancements with time is apparent during our monitoring period. We also notice that the enhancements are coincident with nuclear spirals revealed by the ALMA observation of H30$\alpha$ ~\citep[grey contours]{wang20}, which traces ionized gas in the nuclear mini-spiral.}
\label{fig:sgra_all}
\end{figure}

The results from our LMT/AzTEC monitoring programs in 2014 (ES2) and 2015 (ES3) are shown in Figure~\ref{fig:lcurve_all}. Flux variations appeared on all time scales to which the data are sensitive. Intense flares, which occur on average once a day, were not detected in these $\sim 2$ hr observations, consistent with the finding from longer and overlapping Chandra/JVLA observations. A major long outburst was convincingly detected on May 9, 2014; unfortunately, to our knowledge, there is no radio, near-IR, or X-ray observations covering the same period. Except for this outburst, the amplitudes of the variations ($25\%$) are considerably smaller than what was typically claimed in near-IR or X-ray. All these observations were taken in 2014 just before the pericenter passage of G2, except for two 12 minutes exposures on June 7 and 8, 2015. These latter two show a flux that is consistent with the average of the observations before the passage. 

Figure~\ref{fig:sgra_all} further shows a high S/N image of Sgr A* stacked from all snapshots, with an equivalent exposure time of $\approx 7$ hours. Two enhancements are constantly seen in the North-East and South-East directions of Sgr A* during the monitoring period. These enhancements are coincident with nuclear spirals revealed by the ALMA observation of H30$\alpha$, which traces ionized gas in the nuclear HII region. 

For almost 30 years, Sgr A* has been known to be variable at radio wavelengths ~\citep{brown82}. The amplitude of the intensity modulation at wavelengths between 20 cm and 7 mm \citep{zhao92, zhao01, herrnstein04, falcke04} is 30 to 40\% \citep{macquart06}, with a clear trend that the spectral index $\alpha$ ($S_\nu \propto \nu^{\alpha}$) of the centimeter emission becomes larger when the flux increases \citep{falcke99}. There is also a trend that the amplitude of variability increases at millimeter wavelengths \citep{zhao01, zhao03, miyazaki04, mauerhan05, yusef06}, which can be as large as a factor of a few. The typical time scale of these millimeter fluctuations is about 1.5 to 2.5 hours. A similar variability, both in terms of amplitude and characteristic timescale, is observed at submillimeter wavelengths \citep{eckart06, yusef06, yusef08, yusef09b, marrone08, kunneriath08}. The lightcurve data from this study is consistent with previous works.

\section{Discussion}


\begin{figure*}
\begin{center}
\includegraphics[scale=0.6]{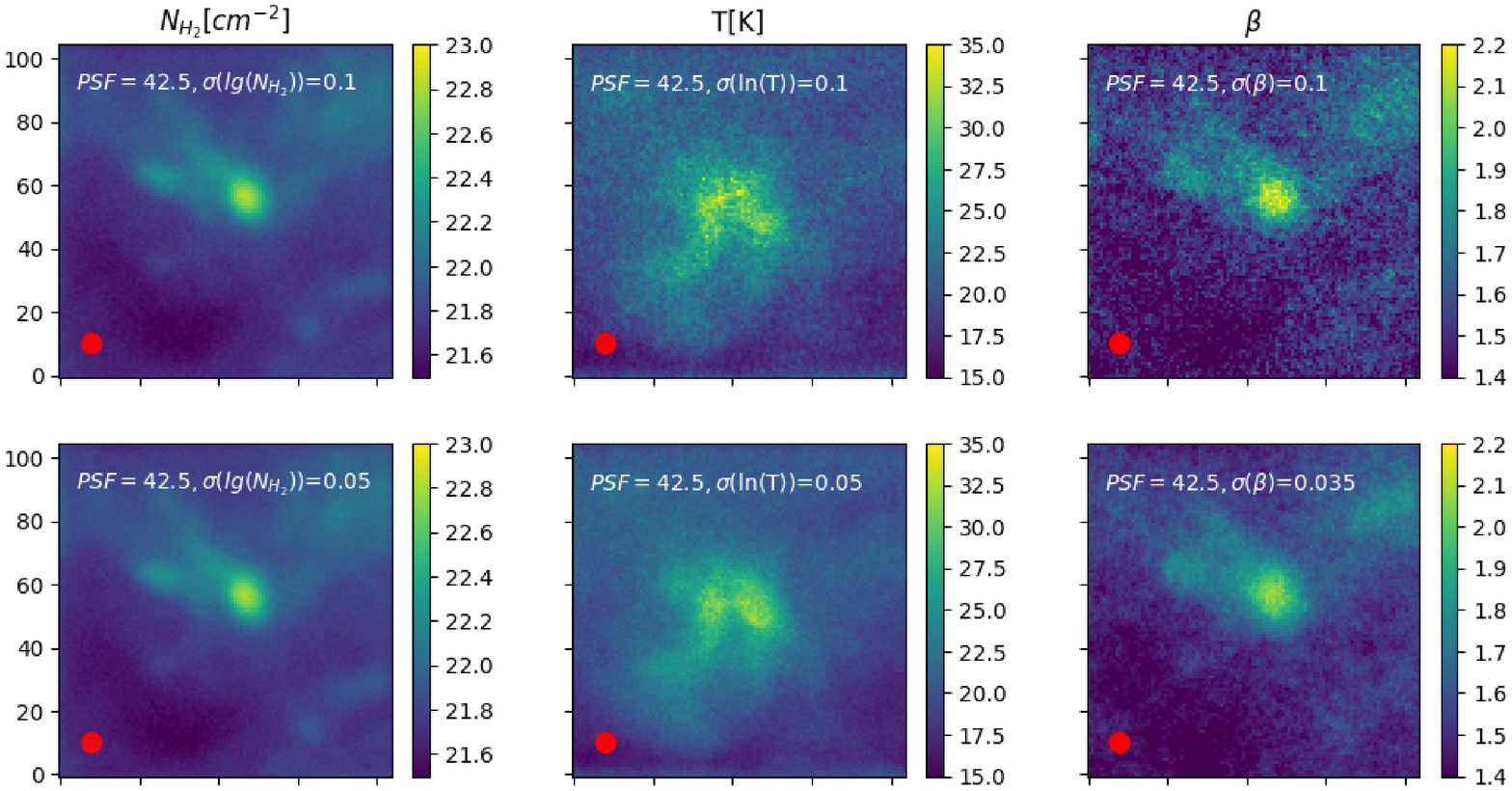}
\includegraphics[scale=0.5]{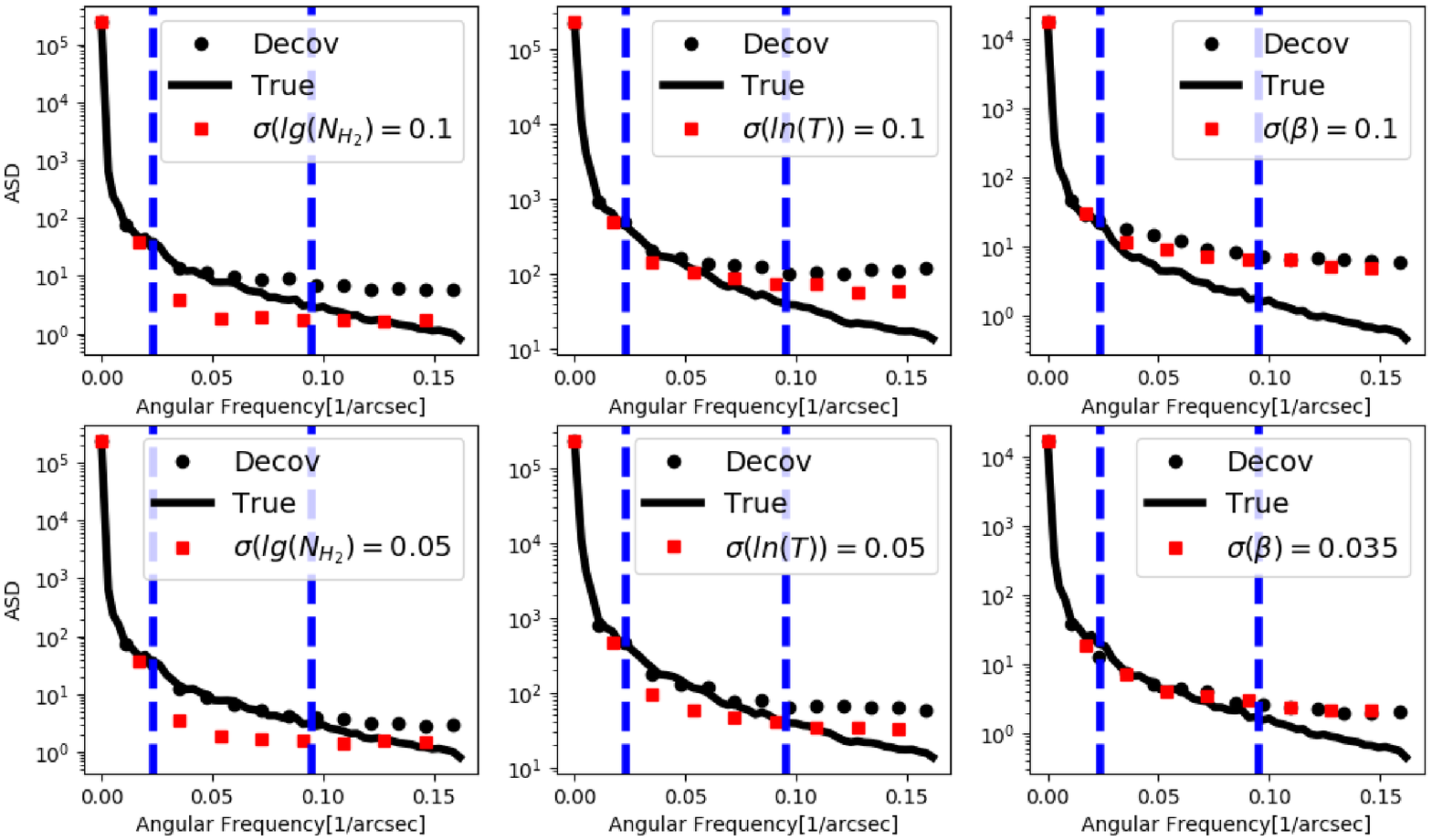}
\end{center}
\caption{Upper: Similar to Figure~\ref{fig:imgs_train}, best-fit maps of $N_{H_2}$, $T$ and $\beta$ for M17 at 8.4 kpc,  derived with the C2LR approach and also using two different smoothness priors. Lower: Same as Figure~\ref{fig:asds_train}, but now including the ASDs of the images plotted above in red squares. The dotted lines show results derived with the MBD approach using the same smoothness prior.}
\label{fig:psf_low}
\end{figure*}

\subsection{Key Questions to be Addressed with a High-Resolution 1.1 mm Map of the CMZ}

Milky Way dust properties have been studied using \textit{Herschel}-only data or \textit{Herschel} plus long-wavelength data, but these studies are mostly limited to nearby Galactic clouds \citep{sadavoy13,sadavoy16,forbrich15}. With a high-resolution 1.1 $mm$ map, via a regularized deconvolution approach, we demonstrate that the CMZ clouds, at a distance of $8.5$ kpc, can be resolved down to sub-pc scales. Regularized deconvolution has been used in astronomy before, mostly using linear models where optimal solutions may be found at a lower computational cost \citep{suyu06, marsh15}. Our Bayesian approach does not require linear models, and can be integrated with other forms of Bayesian analysis, such as Hierarchical Bayesian analysis and Bayes Factor Hypothesis Tests.

The value of $\beta$ and its dependence on $N_{H_2}$ and $T$ in the CMZ differ from what is suggested in nearby clouds, which are observed on spatial scales at least a factor of 10 smaller and have different physical conditions (e.g., lower density, temperature, and turbulence). On the other hand, the methodology adopted also has a non-negligible effect on the determination of $\beta$. The present work provides only baseline data reduction and analysis.  In the follow-up works, this approach will be refined with proper foreground and background subtraction. We will also demonstrate how better understanding of the $\beta$-$T$ relations and the probability distribution function (PDF) of $N_{H_2}$ can be obtained via Hierarchical Bayesian analysis.

\subsection{Model-Based Deconvolution}

In Section~\ref{sec:prior}, we have shown that a significant amount of spatial information is recovered with the MBD approach. The C2LR approach was performed in a conventional way without any prior. Yet, in this latter approach, smoothness priors could also be applied to improve spatial resolutions, which is demonstrated in Figure~\ref{fig:psf_low}. The red squares indicate ASDs of best-fit parameters derived by degrading every image to $42.5''$ resolution, but also being regularized with a smoothness prior, which always offers better performance. It is also shown that, under the same prior, the recoverable amount of spatial information differs from parameter to parameter. For $\beta$, the MBD approach and the C2LR approach produce almost identical ASDs. In contrast, for $N_{H_2}$, improvement in spatial resolution is remarkable with the MBD approach. This difference is easy to understand: while $N_{H_2}$ is predominantly constrained by the high-resolution 1.1 $mm$ map, to determine $\beta$, the 500 $\mu m$ flux is of equal significance. Nevertheless, the high-resolution map constructed with the LMT is a key contributor to effective deconvolution.

\subsection{Future Work}
While the initial AzTEC map demonstrates the potential of the LMT, even when it was still 32 m, the advent of the 50-m LMT will greatly advance the reach of our study of the CMZ, including:

\begin{itemize}
\item A three-color dust map from 1.1-2.0 mm, down to a resolution of $\approx 0.2-0.4$ pc. Recent interferometer observations of the CMZ clouds suggest a characteristic spatial scale of 0.1-1 pc, above which the CMZ clouds are distinct from the ``classic'' Galactic clouds in terms of their smooth density structure \citep{rathborne15}. Below 0.1-1 pc, however, CMZ clouds show similarity to those in the Galactic disk in terms of their core-to star-formation efficiency \citep{lu19}. Therefore, high-resolution dust maps provide critical information on how star-formation and collapse of clouds are halted in the CMZ. 

\item Maps of common dense gas tracers. It is known that the dust temperature in the CMZ is significantly lower than the gas temperature \citep{ginsburg16}. This anomaly is not clearly understood since grain-gas energy exchange is usually considered to be efficient in dense clouds. Turbulence has been proposed as an extra heating source for gas. A detailed comparison between gas temperature and dust temperature could offer further insight into this scenario. Line ratios between different gas tracers can also provide probes of physical conditions from dense clumps to diffuse gas in the CMZ.

\item Line-of-sight locations of clouds and the global stellar light distribution. Both can be simultaneously inferred from fitting a model of a dust column density distribution plus a stellar light distribution model to the multi-band near-IR images (HST, Spitzer, 2MASS, and ground-based surveys).

\item TolTEC, equipped with polarization-sensitive detectors, has been scheduled to survey mm thermal dust polarization in the CMZ and to provide unprecedentedly high-resolution maps of the strong magnetic fields in dense regions. This study will shed light on issues related to both dust properties and the dynamics of molecular clouds. 

\end{itemize}

\section{Summary} \label{sec:summary}

As part of the initial science projects of the 32-m LMT, we have carried out 
large-scale mapping of the CMZ and the monitoring of Sgr A*. These projects have enabled us to test various technical approaches for large-scale mapping and time-domain observations with the LMT. The analysis of our obtained data, in combination or comparison with existing complementary data, have further led the following preliminary results and conclusions:

\begin{itemize}

\item We have presented the AzTEC map with a resolution of 10.5", along with the observation strategy and the data reduction approach used to construct the map. We have demonstrated that, via a model-based deconvolution approach, a joint analysis of the AzTEC map in combination with other lower-resolution maps, including \textit{Herschel}, \textit{Planck}, and CSO maps, provides high-resolution maps of $N_{H_2}$, $T$ and $\beta$ across the entire CMZ.

\item We have presented results from a 1.1 mm continuum monitoring program of Sgr A*, using the LMT/AzTEC, as part of a worldwide, multi-wavelength, monitoring campaign of the central massive black hole of our Galaxy. This program is timely because it was conducted at the pericenter epoch of the G2 object, which was expected to be torn apart by the black hole and to contribute to its accretion flow, a scenario which comes under debate recently~\citep{plewa17}. This program with a total exposure of about 9 hours consisted of two components: (1) daily monitoring when the weather permitted; (2) intraday variability study during two coordinated Chandra/JVLA observations of Sgr A*. We detect no significant systematic 1.1 mm flux change of Sgr A* during the program or compared with previous observations other than a $\approx 40\%$ outburst. In fact, the source appears to be remarkably steady, with its flux variation less than about 10\% (rms). We also detect no significant flares during the two intraday studies, as confirmed by the simultaneous X-ray and radio observations. These results have provided constraints on the interplay of the G2 object with the accreting material during the passage.

\end{itemize}

\section*{Acknowledgement}

The AzTEC instrument was built and operated through support from NSF grant 0504852 to the Five College Radio Astronomy Observatory. The authors gratefully acknowledge the many contributions of David Hughes in leading the LMT to its successful operational state. This work is partly supported by NASA via the grant NNX17AL67G. This work is also supported by Consejo Nacional de Ciencia y Tecnolog\'{\i}a (CONACYT) M\'{e}xico research grant CB-A1-S-28458. This work is based on observations made with \textit{Herschel}, \textit{Planck} and the Caltech Submillimeter Observatory (CSO) telescope. \textit{Herschel} is a European Space Agency cornerstone mission with science instruments provided by European-led Principal Investigator consortia and with significant participation by NASA. \textit{Planck} is a project of the European Space Agency with instruments funded by ESA member states, and with special contributions from Denmark and NASA. CSO was operated by the California Institute of Technology under cooperative agreement with the National Science Foundatio.

\section*{Data Availability}
The AzTEC data and products underlying this article are available at \footnote{\url{https://github.com/tangyping/products.git}}. The \textit{Herschel} Hi-Gal products were provided by the Hi-Gal team by permission. The datasets from \textit{Planck} telescope were derived from \footnote{\url{https://pla.esac.esa.int/}}. The datasets from CSO/Bolocam were derived from \footnote{\url{https://irsa.ipac.caltech.edu/data/BOLOCAM_GPS}}.

\appendix

\section{Slice Sampling} \label{apendix:slice}

In statics, Gibbs sampling is a sampling strategy to obtain samples from the joint distribution of a posterior $P(x_1, x_2, x_3...)$ by sampling from the full conditional posterior of each parameter $x_j$ in turn. This means, if we start from the $i$th sample $(x_1^i, x_2^i, x_3^i...)$ of a posterior distribution $P(x_1, x_2, x_3...)$, the $i+1$th sample $(x_1^{i+1}, x_2^{i+1}, x_3^{i+1}...)$ could be obtained by sampling from each conditional posterior $P(x_i | x_1^i, x_2^i, ... x_{k-1}^i, x_{k+1}^i, x_{k+2}^i...)$ in turn. In this way, sampling from a multi-dimensional posterior distribution is reduced to sampling from multiple 1-D conditional posterior distributions. 

\begin{figure*}
\includegraphics[scale=0.55]{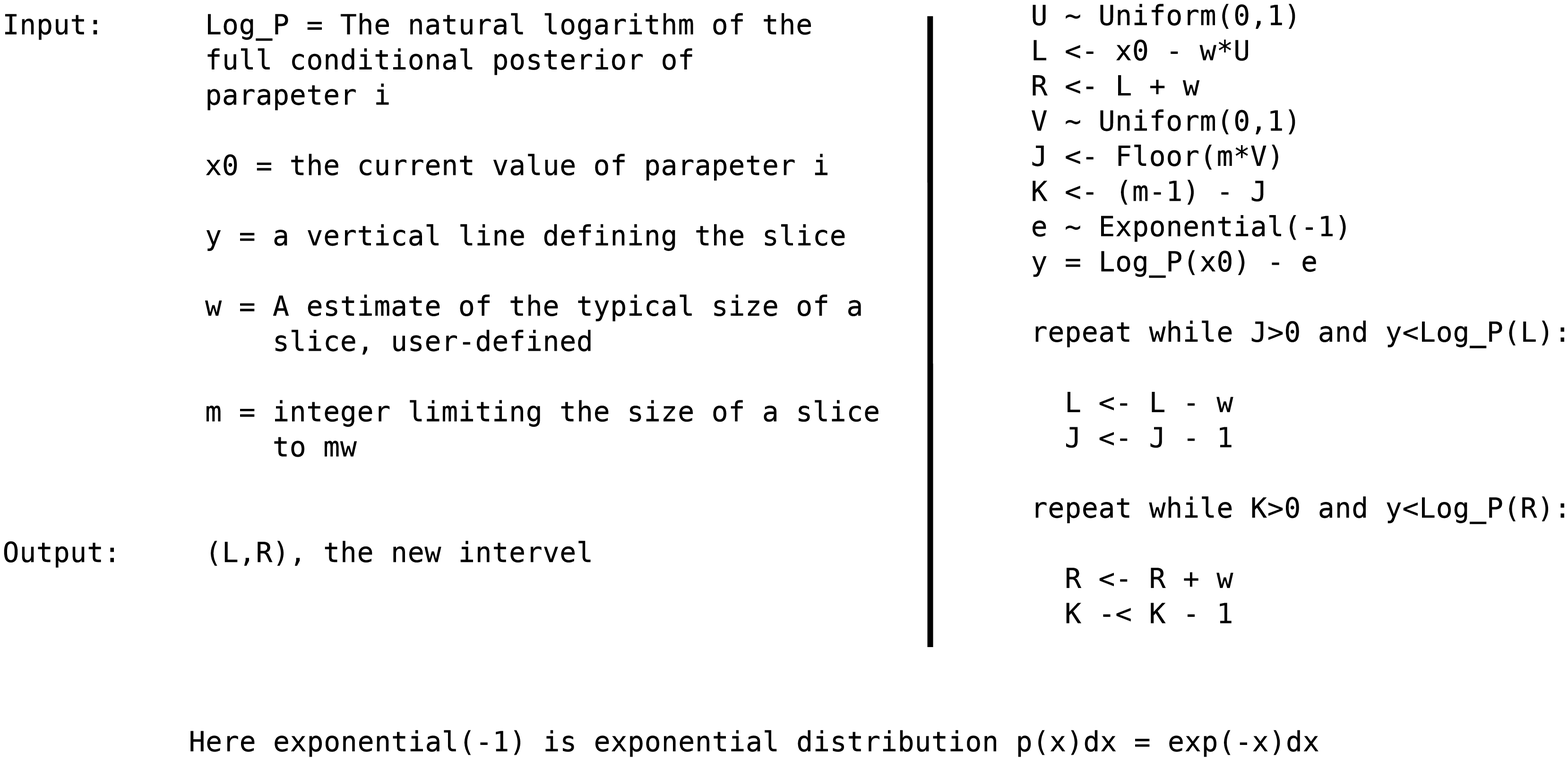}
\caption{The stepping-out procedure of slice sampling, adapted from \cite{neal03}.}
\label{fig:slice_step}
\end{figure*}

\begin{figure*}
\includegraphics[scale=0.55]{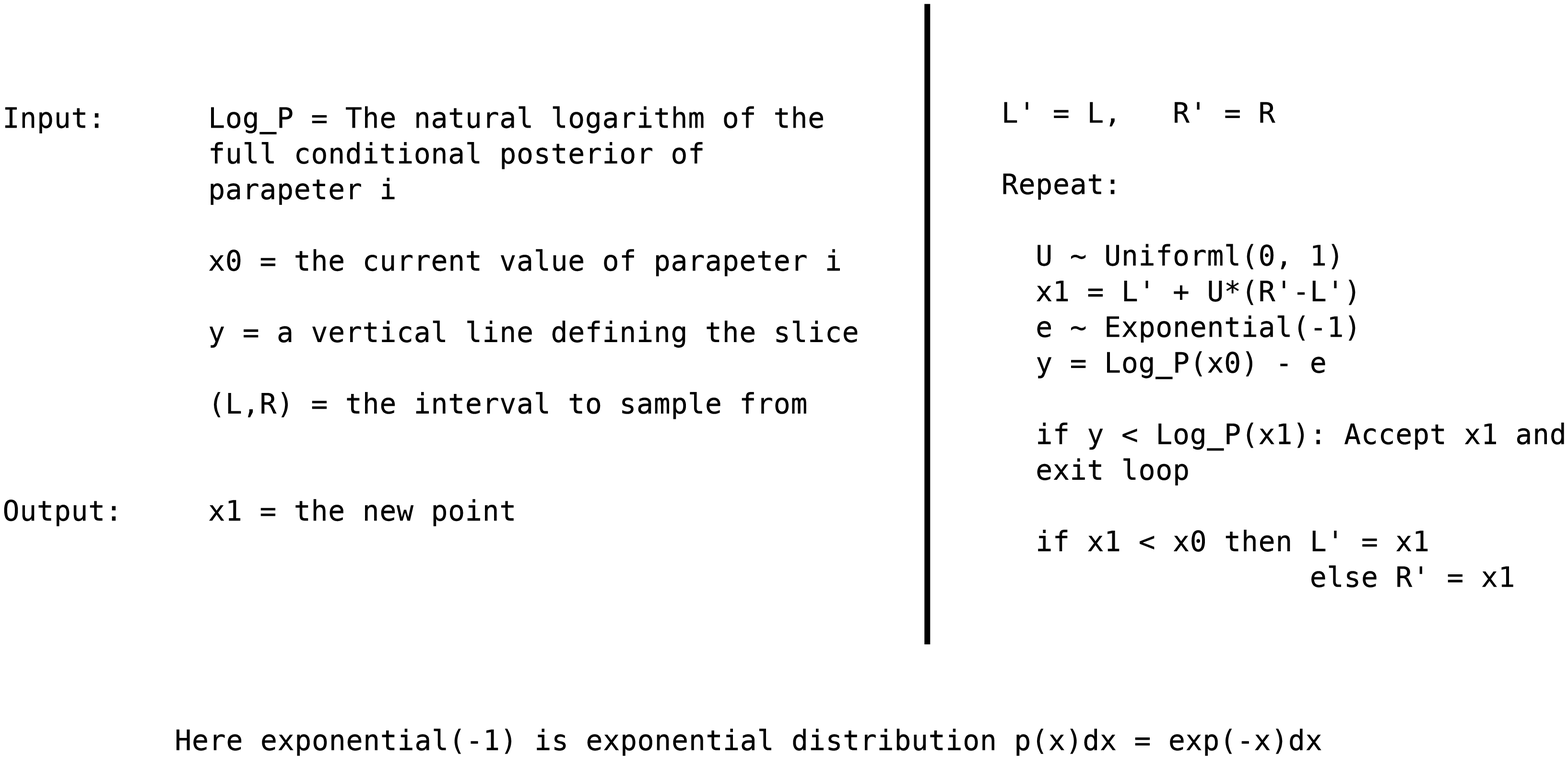}
\caption{The shrinkage procedure in slice sampling, adapted from \cite{neal03}}
\label{fig:slice_shrink}
\end{figure*}

Following \cite{neal03}, we further use slice sampling with a stepping out and shrinkage strategy to sample from the full conditional posterior of each parameter ($N_{H_2,ix,iy}/T_{ix,iy}/\beta_{ix,iy}$) in a grid cell. This procedure is composed of two steps, stepping out, which is illustrated by Figure~\ref{fig:slice_step}, and shrinkage, which is illustrated by Figure~\ref{fig:slice_shrink}). In practice, the stepping out procedure could be omitted by directly shrinking from the largest interval (e.g., lower limits and upper limits). This is equivalent to setting $w$ Figure~\ref{fig:slice_step} to an infinitely large number. We find that this later approach has acceptable efficiency for our model.

\begin{figure}
\begin{center}
\includegraphics[scale=0.5]{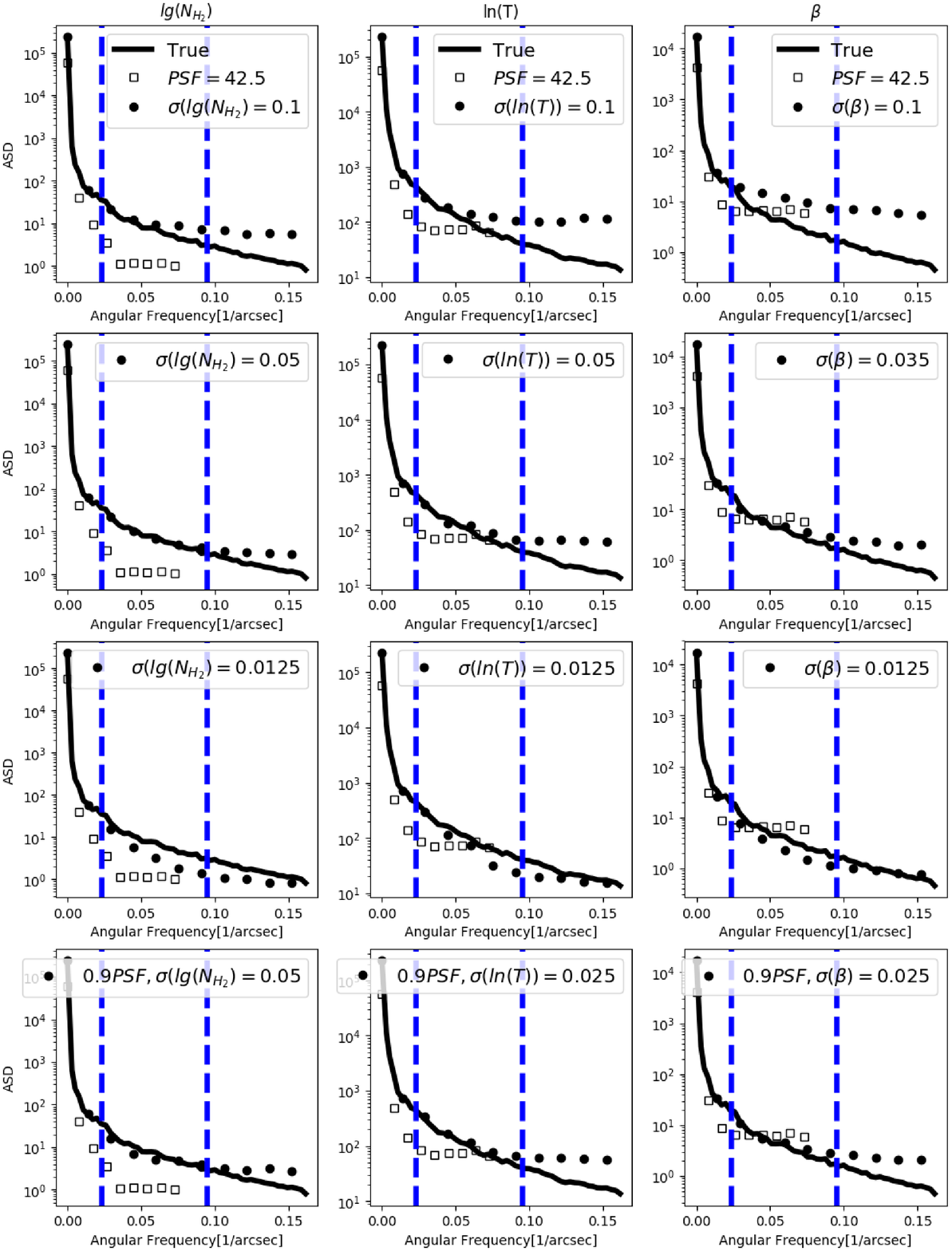}
\end{center}
\caption{The ASDs of each image plotted in Figure~\ref{fig:imgs_train}. The frequencies are scaled by a factor of $\frac{\pi}{2ln(2)}$, similar to Figure~\ref{fig:asd_maps}. The dotted lines indicate best-fit results using different smoothness priors. The ``true'' ASD and the ASD derived with the C2LR approach are also shown in solid lines and squares as references. The last row shows ASDs derived with underestimated PSFs ($0.9 \times FWHM_{true}$). In each plot, the vertical line on the left indicates the largest PSF ($FWHM=42.5''$ at $500$ $\mu m$), and the vertical line on the right indicates the smallest PSF ($FWHM=10.5''$ at 1.1 mm). Increasing strengths of a smoothness prior (smaller $\sigma$) result in lower noise but also lower amplitudes on intermediate to large spatial scales.}
\label{fig:asds_train}
\end{figure}

The log-likelihood $ln\_P(x)$ for $x \in (T_{ix,iy}, N_{H_2,ix,iy}, \beta_{ix,iy})$ is given by:

\begin{equation}
ln\_P(x) \propto \sum_{j,ix,iy} \frac{(Mod(x, ix, iy, \nu_j) - Map(ix,iy,\nu_j))^2}{2 \times \sigma^2_{ix,iy,\nu_j}}
\end{equation}

Where j refers to the $j$th band, $\mathbf{Mod(\mathbf{x}, \nu_j)} = \mathbf{F(\mathbf{x}, \nu_j)} \otimes \mathbf{beam_j}$. In practice, one can avoid a full convolution of $\mathbf{F(\nu_j)}$ with $\mathbf{beam_j}$ every time a single parameter is updated. Instead, if the current model map $\mathbf{Mod(x_{i}, \nu_j)}$ is recorded, $\mathbf{Mod(x_{i+1}, \nu_j)}$ can be calculated by $\mathbf{Mod(x_{i+1}, \nu_j)} = \mathbf{Mod(x_{i}, \nu_j)} + (F(x_{i+1}, \nu_j) - F(x_{i}, \nu_j) \times \mathbf{beam_j}$. In this study, we choose to truncate $\mathbf{beam_j}$ at a radius of $4\sigma$.

\section{Testing Different Smoothness Priors} \label{apendix:prior}

To examine how different choices of $\sigma$ (or $\lambda$ in Eq~\ref{eq:smprior}) affect our SED analysis, we first construct ``true'' maps of $N_{H_2}$ and $T$ in M17, which are best-fit maps derived by degrading the Hershel $160-500$ $\mu m$ maps to a common resolution of FWHM = $42.5''$, assuming $\beta=2.0$. \textit{Herschel} PACS (Level 2.5) and SPIRE (Level 3) product maps are downloaded from the ESA \textit{Herschel} Science Archive. The observations are obtained from the HOBOYS survey \citep{motte10} and the Hi-Gal survey. The peak surface brightness of M17 at $500$ $\mu m$ is $>5000$ mJy/sr, similar to that of major CMZ clouds. At 2.1 kpc, these ``true'' maps are resolved to a physical scale approximately equal to our highest resolution CMZ map at 1.1 $mm$ (i.e., the highest resolution achievable by any deconvolution technique). In order to also generate a ``true'' map of $\beta$, we assume a positive relationship between $\beta$ and $lg(N_{H_2})$:

\begin{equation}
\beta = A (\frac{lg(N_{H_2)}}{lgN_t})^{-\alpha_1} \{ \frac{1}{2} [ 1 + (\frac{lg(N_{H_2})}{lgN_t})^{\frac{1}{\delta}} ] \}^{(\alpha_1 -\alpha_2)\delta} 
\end{equation}

Here, $A=0.72$, $lgN_t=22.71$, $\alpha_1=-0.20$ and $\alpha_2=-0.02$. The relationship is derived from an analysis of $N_{H_2}-\beta$ relation for CMZ clouds, which will be present in a separate paper focusing on dust properties. 

The $160$ $\mu m$ to 1.1 mm SEDs of M17, assuming it is at the distance of the CMZ ($\approx8.4$ kpc), are then simulated using the ``true'' maps of $\{N_{H_2},T,\beta \}$ generated above. After this, we perform our regularized Bayesian analysis on the simulated SEDs with different strengths of smoothness prior: 
\\
\\
a) a weak prior, with $\sigma_{lg(N_{H_2})}=0.1, \sigma_{ln(T)}=0.1, \sigma_{\beta}=0.1$; 
\\
\\
b) a moderate prior, with $\sigma_{lg(N_{H_2})}=0.05, \sigma_{ln(T)}=0.05, \sigma_{\beta}=0.035$; 
\\
\\
c) a strong prior, with $\sigma_{lg(N_{H_2})}=0.0125, \sigma_{ln(T)}=0.0125, \sigma_{\beta}=0.0125$.
\\
\\
Furthermore, to examine whether this analysis is sensitively dependent on the correctness of the PSF models, we test a case in which $160$ $\mu m$-1.1 mm PSF sizes are all underestimated by $10\%$, with $\sigma$ values identical to those for prior b).

The 1-D ASDs of each best-fit image in Figure~\ref{fig:imgs_train} are shown by dotted lines in Figure~\ref{fig:asds_train}, where the ASDs of the ``true'' maps and the ASDs of the best-fit maps derived from the C2LR approach are plotted for comparison. Apparently, the performance of the MBD approach is superior to the C2LR approach within a broad range of $\sigma$, from $\sigma(lg(N_{H_2}[cm^{-2}), ln(T[K]), \beta)=0.0125-0.1$. The C2LR approach loses a significant amount of spatial information below the sizes of the largest PSF (FWHM = $42.5''$), which is at least partially recoverable, depending on the choice of $\sigma$. 

These simulations demonstrate that the strength of the smoothness prior ($\sigma$) is, to some degree, self-assessable, even when a training dataset (observations of M17 in our case) is not available. If the smoothness prior is too weak, all best-fit ASDs show high levels of white noise. A flat ASD on scales much larger than the smallest PSF could be an indication of overfitting to noise, due to a lack of constraining power from smoothness priors. As the strength of the smoothness prior increases, the ASDs start to fall below a threshold level set by the C2LR approach. If this occurs at angular sizes larger than the largest PSF, which is an extreme case and is not seen even with our strongest priors, the smoothness prior is likely over-weighted.

\end{document}